\documentclass[
superscriptaddress,
amsmath,amssymb,
pra,
floatfix,twocolumn
]{revtex4-1}

\usepackage{verbatim}
\usepackage{graphicx}
\usepackage{dcolumn}
\usepackage{bm}
\usepackage{hyperref}
\usepackage{physics}
\usepackage[utf8]{inputenc}

\usepackage{color}

\def\equationautorefname~#1\null{%
  Eq.~(#1)\null
}

\def\figureautorefname~#1\null{%
  Fig.~#1\null
}

\begin{document}

\title[Sample title]{Weak localization corrections to the thermal conductivity in \texorpdfstring{$s$}{s}%
-wave superconductors}

\author{L. Gonz\'alez Rosado}
\affiliation{JARA Institute  for  Quantum  Information,  RWTH  Aachen  University,  52056  Aachen, Germany}
\affiliation{JARA Institute for Quantum Information (PGI-11), Forschungszentrum J\"ulich, 52425 J\"ulich, Germany}

\author{F. Hassler}
\affiliation{JARA Institute  for  Quantum  Information,  RWTH  Aachen  University,  52056  Aachen, Germany}
\author{G. Catelani}
\affiliation{JARA Institute for Quantum Information (PGI-11), Forschungszentrum J\"ulich, 52425 J\"ulich, Germany}

\date{February 2020}

\begin{abstract}
We study the thermal conductivity in disordered $s$-wave superconductors.  Expanding on previous works for normal metals, we develop a formalism that tackles particle diffusion as well as the weak localization (WL) and weak anti-localization (WAL)
effects. Using a Green's functions diagrammatic technique, which takes into account the superconducting nature of the system by working in Nambu space, we identify the system's low-energy modes, the diffuson and the Cooperon. The time scales that characterize the diffusive regime are energy dependent; this is in contrast with the the normal state, where the  relevant time scale is the mean free time $\tau_e$, independent of energy.
The energy dependence introduces a novel energy scale $\varepsilon_*$, which in disordered superconductors ($\tau_e \Delta\ll 1$, with $\Delta$ the gap) is given by $\varepsilon_* = \sqrt{\Delta/\tau_e}$.
From the diffusive behavior of the low-energy modes, we obtain the WL correction to the thermal conductivity. We give explicitly expressions in two dimensions.
We determine the regimes in which the correction depends explicitly on $\varepsilon_*$ and propose an optimal regime to verify our results in an experiment.
\end{abstract}
\maketitle

\section{\label{sec:intro}INTRODUCTION}

The study of quantum effects in the transport properties of disordered conductors has a long history. For thermal conductivity in normal metals, a fundamental question was whether such corrections obey the Wiedemann-Franz (WF) law relating the electrical conductivity $\sigma$ to thermal conductivity $K$~\cite{WF}. For non-interacting electrons, the WF law is expected to hold with the inclusion of quantum corrections in the weak localization regime but the numerical coefficient known as the Lorentz number $L_0=K/\sigma T$, with $T$ the temperature, is reduced when approaching the Anderson localization transition~\cite{Enderby1994}. Away from the transition, deviations have been calculated due to electron-electron interactions~\cite{Catelani2005}. Mesoscopic fluctuations can also lead to violations of the WF law~\cite{Vavilov2005}. In the superconducting state the dc electrical resistance vanishes and hence there is no WF law; in fact, approaching the critical temperature from the normal state, superconducting fluctuations lead to a divergent electrical conductivity, whereas they only constitute a finite correction to $K$~\cite{Niven2002}. Sufficiently far below the critical temperature, fluctuations are negligible and the leading order expression for the thermal conductivity of a BCS superconductor has been obtained in the early work of Ref.~\cite{Bardeen1959}. Further extensions to this result include the effects of electron-phonon scattering~\cite{KM1961,Tewordt1962}, strong coupling~\cite{Ambegaokar}, and paramagnetic impurities~\cite{AmbGrif1965}. However, to the best of our knowledge, the question of the fate of the weak localization correction to the thermal conductivity in the superconducting state has so far only been addresses for SNS junctions~\cite{HHS} and not for the bulk.

In this paper, we analytically calculate  
the weak localization correction to the thermal conductivity in $s$-wave superconductors, including weak anti-localization in a system with spin-orbit scattering. 
To that end, we extend the formalism used to study diffusion in normal metals, see e.g. Ref.~\cite{AM}, so that it can be used for superconductors as well. Technically, we work with matrix Green's functions in Nambu space.
In the next section, we introduce the model for disordered superconductors to establish our notation.
In \autoref{sec:disorder}, we study
diffusion in disordered superconductors in depth by generalizing the ladder approximation.
We focus on the A-type diffusons and Cooperons~\cite{AZ}, since in a time reversal invariant system, the D-type diffusons do not contribute to thermal transport~\cite{Ambegaokar}.
In contrast to the normal state, the diffusion constant in the superconducting state depends on energy (measured from the Fermi energy). This energy dependence manifests itself in the condition defining the diffusive regime in the time domain, which is now not simply given by the requirement of time being long compared to the impurity scattering time $\tau_e$. We find that the corresponding time scale in the superconducting state is different for energies below or above an energy scale $\varepsilon_*$ which is a function of the superconducting gap $\Delta$ and the scattering time; for disordered superconductors with $\tau_e \Delta \ll 1$, we find $\varepsilon_* = \sqrt{\Delta/\tau_e}$.

In \autoref{sec:thermalconductivity}, we make use of the results of the preceding section to calculate the thermal conductivity from the Kubo formula. We recover previous results~\cite{Bardeen1959,Ambegaokar} for the Drude-Boltzmann contribution to the thermal conductivity, which, because of the opening of the superconducting gap, is suppressed as temperature is reduced. As the diffusion constant is energy dependent, we have to specify whether the phase-coherence length or the phase-coherence time is constant in a material in order to evaluate the weak localization correction. We obtain results for both scenarios;
in general, the WL correction is temperature dependent. Interestingly, the suppression of the WL correction with decreasing temperature is generally stronger than that of the Drude-Boltzmann term. Of possible experimental interest is the temperature region of order $T_\Delta \approx 0.9T_c$ defined by $k_BT_{\Delta}=\Delta(T)$. On one hand, this temperature is sufficiently high that the strong (exponential) suppression of the (Drude-Boltzmann) thermal conductivity has not yet taken place. On the other hand, for disordered superconductors, this temperature is low enough that most of the weak localization correction is already suppressed. This temperature is therefore optimal in order to observe the deviation of the WL correction in the superconducting state from its normal-state value, as we predict the thermal conductivity to be larger than expected from its value just above $T_c$.
We summarize our findings in Sec.~\ref{sec:conclusiones}. A number of details can be found in Appendices~\ref{appendix:particlediffusion} to \ref{appendix:so}.

\section{\label{sec:model}Model}

The (mean field) Hamiltonian for a superconductor with $s$-wave pairing can be expressed in the Bogoliubov-de Gennes (BdG) form as~\cite{BdG}
\begin{equation}
H=\sum_{\bm{k}}\bm{\Psi}_{\bm{k}}^\dag \hat{H}_{\text{BdG}}(k) \bm{\Psi}_{\bm{k}}.
\label{eq:Hbcs}
\end{equation}%
with the Nambu vector
\begin{equation}\bm{\Psi}_{\bm{k}}= \begin{pmatrix} c_{\bm{k}\uparrow} \\ c^\dag_{-\bm{k} \downarrow} \end{pmatrix},
\label{eq:nambubasis}\end{equation}
where $c^\dag_{\bm{k} \sigma}$ and $c_{\bm{k} \sigma}$ are creation and annihilation operators for electrons with momentum $\bm{k}$ and spin $\sigma$, respectively. The BdG Hamiltonian is given by
\begin{equation}
\hat{H}_{\text{BdG}}(k)=\epsilon_k \tau_3 - \Delta \tau_1,
\label{eq:Hbdg}
\end{equation}%
where the hat denotes matrices in the Nambu space. Here, $\epsilon_k=  k^2/2m-\mu$, $m$ is the electron mass, $\mu= k_F^2/2m$ the Fermi energy with $k_F$ the Fermi momentum, and $\tau_i$ the Pauli matrices in Nambu space (we omit hats on these matrices for notational simplicity).
For later use, we introduce the basis $\{\ket{\text{e}},\ket{\text{h}}\}$ in Nambu space, where the states $\ket{\text{e}}$ and $\ket{\text{h}}$ stand for electron and hole respectively.
The Bogoliubov-de Gennes Hamiltonian (\ref{eq:Hbdg}) includes the non-interacting electron and hole Hamiltonians in its diagonal terms as well as the
pairing term, given by the superconducting order parameter $\Delta$,
 in its off-diagonal terms. The retarded and advanced Green's functions are then solutions of
\begin{equation}
(E - \hat{H}_{\text{BdG}}  \pm i0^+)\hat{G}_{E}^{R,A}=1.
\label{eq:greeneq}
\end{equation}%
 We distinguish the four different elements of the matrix Green's function as follows
\begin{equation}
    \hat{G}_E^{R,A}=\begin{pmatrix}G_E^{R,A} & F_E^{R,A}\\ \bar{F}_E^{R,A} & \bar{G}_E^{R,A} \end{pmatrix}.
    \label{eq:greennambu}
\end{equation}
The diagonal terms---that is, the electron and hole Green's functions---describe electron and hole propagation, respectively. The off-diagonal terms, known as anomalous Green's functions, account for particle-hole conversion, \textit{i.e.}, Andreev reflection.

So far, we have considered a clean superconductor. To treat the elastic scattering of electrons off impurities we introduce a random disorder potential.
The disorder potential $\hat{V}(\bm{r}) = V(r) \tau_3$ is taken to be Gaussian distributed 
with $\overline{V(r)}=0$, where the overline $\overline{\cdots}$ denotes the disorder average.  We work in the weak disorder limit $k_F l_e \gg 1$, where $l_e$ is the mean free path, which allows for the perturbative treatment of impurity scattering. We define the disorder parameter $\gamma_e$ by relating it to the disorder average of the variance of the potential such that
\begin{equation}
    \overline{\hat{V}(\bm{r})\otimes\hat{V}(\bm{r}')}=\gamma_e \delta^{(d)}(\bm{r}-\bm{r}')\hat{U}_v,
    \label{eq:disorder}
\end{equation}%
where $\hat{U}_v=\tau_3 \otimes \tau_3$.
The disorder parameter is related to the scattering time $\tau_e=l_e/v_F$ in the normal state and to the normal-state density of states per spin $\rho_0$ as $\gamma_e= 1/2\pi\rho_0 \tau_e$ with $v_F= k_F/m$ the Fermi velocity.

In a normal metal, both electrical and thermal conductivity are attributed to free conduction electrons, and both phenomena can be understood by studying electron diffusion. In superconductors, the (super)current is carried by Cooper pairs;
the thermal conductivity, however, is still related to particle diffusion. In order to study diffusion in the superconducting state, in the next section
we develop a matrix formalism in Nambu space that
enables us to generalize the diagrammatic approach well established in the study of diffusion in the normal state.

\section{\label{sec:disorder}Particle diffusion and weak localization}

\begin{figure}[tb]
  \includegraphics[width=\linewidth]{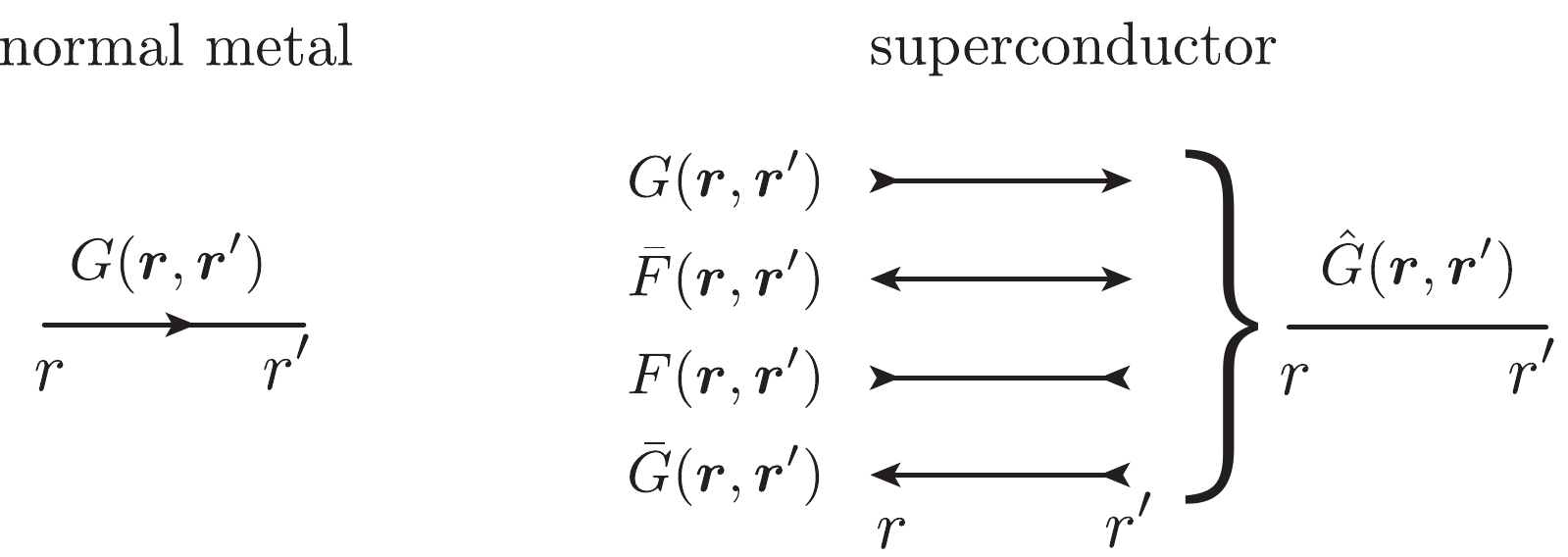}
    \caption{Feynman diagrams for the time ordered Green's functions in a normal metal and a superconductor. Time evolution occurs from left to right. The four components of the Green's function for a superconductor are distinguished by the arrows at the ends. The matrix formulation in the Nambu formalism is represented by an arrow-less line.}
    \label{fig:greens}
\end{figure}

\begin{figure}[tb]
  \includegraphics[width=\linewidth]{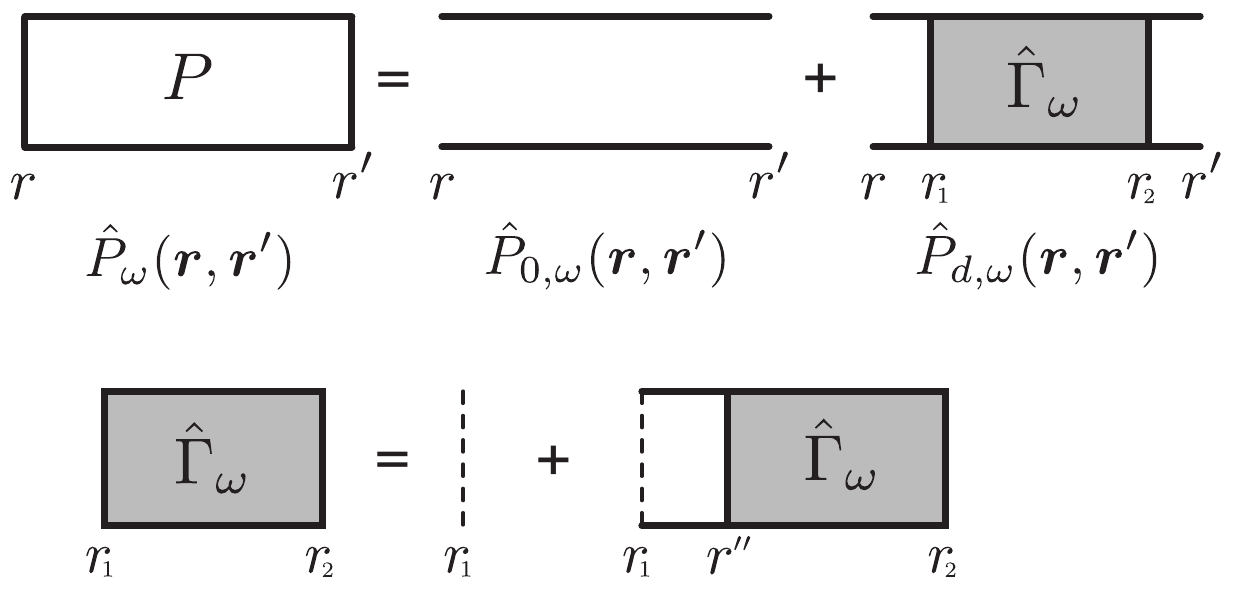}
    \caption{Representation of the ladder approximation for the diffusion probability matrix $\hat{P}_{\omega}(\bm{r},\bm{r}')$ in a superconductor as defined in Eq.~(\ref{eq:pwdef}). The upper arrow-less line represents the disorder-averaged retarded superconducting Green's function expressed in matrix form in the Nambu formalism. The lower line represents its complex conjugate. The dashed lines represent impurity scattering. The total probability of diffusion $\hat{P}_{\omega}(\bm{r},\bm{r}')$ is composed by the Drude-Boltzmann contribution $\hat{P}_{0,\omega}(\bm{r},\bm{r}')$ and the diffuson $\hat{P}_{d,\omega}(\bm{r},\bm{r}')$. The latter includes the structure factor $\hat{\Gamma}_{\omega}(\bm{r}_1,\bm{r}_2)$ which accounts for elastic scattering with static impurities.}
    \label{fig:ladder}
\end{figure}

In this section, we study the propagation of particles  in disordered conventional superconductors in the weak disorder limit $k_F l_e \gg 1$. In this limit, localization affects the transport coefficients, but Anderson localization~\cite{Anderson} does not yet take place. %
Throughout this section, we expand to the superconducting state the diagrammatic treatment of particle propagation in a normal metal presented in chapter 4 of Ref.~\cite{AM}.
The main technical change involves modifying the Feynman diagrams to include all the four components of the superconducting Green's function defined in \autoref{eq:greennambu}~\cite{AGD}, see \autoref{fig:greens}.
We define the quantum diffusion probability matrix $\hat{P}_{\omega}(\bm{r},\bm{r}')$ as
\begin{equation}
\hat{P}_\omega(\bm{r},\bm{r}')=\overline{\hat{G}^R_{E+ \omega}(\bm{r},\bm{r}')\otimes \hat{G}^A_E(\bm{r}',\bm{r})^T},
\label{eq:pwdef}
\end{equation}%
where the retarded Green's functions in real space are given by $\hat{G}^{R}_{E}(\bm{r},\bm{r}')=\bra{\bm{r}'}\hat{G}^{R}_{E}\ket{\bm{r}}$ and
\begin{equation}
    \hat{G}^A_E(\bm{r}',\bm{r})^T=\hat{G}^R_{E}(\bm{r},\bm{r}')^*.
    \label{eq:greenra}
\end{equation}
The matrix $\hat{P}_\omega(\bm{r},\bm{r}')$ acts on the space spanned by $\ket{i,j}=\ket{i}\otimes\ket{j}$ with $i,\, j\in\{ \mathrm{e},\,\mathrm{h}\}$; that is, $\ket{i}$ and $\ket{j}$ are basis states in the Nambu spaces pertaining to the retarded and advanced Green's functions, respectively.
We discuss the proper normalization of this probability in Appendix~\ref{appendix:particlediffusion}. We stress that $\hat{P}_{\omega}(\bm{r},\bm{r}')$ depends on the energy argument $E$ appearing in the Green's functions, although we do not highlight this in the notation for simplicity: scattering off impurities being elastic, the energy argument can be treated as a parameter that is constant during diffusion. The diagrammatic expression for $\hat{P}_{\omega}(\bm{r},\bm{r}')$ in the ladder approximation is shown in \autoref{fig:ladder}. In each diagram shown in the figure, the upper line represents the retarded Green's function in Nambu space from point $\bm{r}$ to point $\bm{r}'$, and the lower one represents its complex conjugate, given by \autoref{eq:greenra}.%
We calculate three main contributions to particle propagation, starting with the Drude-Boltzmann contribution $\hat{P}_{0,\omega}(\bm{r},\bm{r}')$. This contribution accounts for the probability of propagation in a disordered medium without scattering off any impurity. Subsequently, we include classical scattering events and calculate the diffuson $\hat{P}_{d,\omega}(\bm{r},\bm{r}')$. We show that in the superconducting state the so-called diffusive or hydrodynamic approximation is applicable beyond a time scale that differs from that of the normal state and depends on energy $E$. We define the total probability of diffusion as the sum of these two contributions%
\begin{equation}
    \hat{P}_{\omega}(\bm{r},\bm{r}')=\hat{P}_{0,\omega}(\bm{r},\bm{r}')+\hat{P}_{d,\omega}(\bm{r},\bm{r}').
    \label{eq:pwterms}
\end{equation}%

In the last part of the section, we consider the effect of coherent backscattering and derive the weak localization correction to particle diffusion $\hat{P}_{c,\omega}(\bm{r},\bm{r}')$, that is the Cooperon contribution. In this way we generalize previous studies of weak localization in superconductors, which considered the effect on the density of superconducting electrons~\cite{SA} and on non-local transport in normal/superconductor/normal structures~\cite{DM,DM2}.

\subsection{Drude-Boltzmann contribution}

The Drude-Boltzmann contribution $\hat{P}_{0,\omega}(\bm{r},\bm{r}')$ is given by
\begin{equation}\hat{P}_{0,\omega}(\bm{r},\bm{r}')=\overline{\hat{G}^{R}_{E+ \omega}(\bm{r},\bm{r}')} \otimes \overline{\hat{G}^A_E(\bm{r}',\bm{r})^T}.
\label{eq:p0def}\end{equation}%
The disorder-averaged superconducting retarded Green's function can be explicitly calculated in momentum space, where it is given by~\cite{AGD}%
\begin{equation}
\overline{\hat{G}^{R}_E(k)}=\frac{\bar{E}\tau_0+\epsilon_k\tau_3-\bar{\Delta}\tau_1}{\bar{E}^2-\epsilon_k^2-\bar{\Delta}^2}.
\label{eq:greenk}
\end{equation}%
with
\begin{equation}\bar{E}=E\left[1+i \frac{ 1}{2 \tau_e} \frac{\text{sgn}(E)}{\sqrt{E^2-\Delta^2}}\right],
\label{eq:ebar}\end{equation}%
and
\begin{equation}\bar{\Delta}=\Delta\left[1+ i \frac{ 1}{2 \tau_e}\frac{ \text{sgn}(E)}{ \sqrt{E^2-\Delta^2}}\right].
\label{eq:dbar}\end{equation}%
The Fourier transform of \autoref{eq:greenk} into real space can then be calculated in the limit $\mu \gg \epsilon, \Delta$, with $\epsilon=\sqrt{E^2-\Delta^2}$, by linearizing the spectrum around $k_F$. We provide the explicit result in the two dimensional case, since it will be of particular interest for the weak localization correction. In the limit $k_F R \gg 1$ with $\bm{R}=\bm{r}'-\bm{r}$, we have ($E>0$)
\begin{equation}
\begin{split}
\overline{\hat{G}^{R}_E(\bm{r},\bm{r}')}=\frac{m}{\sqrt{2 \pi k_F R }} e^{i\frac{ \epsilon R}{ v_F }-\frac{R}{2l_e}} \bigg[i \frac{E}{\epsilon}\cos(k_F R+\frac{3\pi}{4})\tau_0  \\ - \sin(k_F R+\frac{3\pi}{4})\tau_3  - i \frac{\Delta}{\epsilon}\cos(k_FR+\frac{3\pi}{4})\tau_1 \bigg].
\end{split}
\label{eq:greenr}
\end{equation}%
The advanced Green's function in real space can be arrived at using \autoref{eq:greenra}.
Having obtained the disorder-av\-eraged superconducting Green's functions in real space, $\hat{P}_{0,\omega}(\bm{r},\bm{r}')$ can be found from \autoref{eq:p0def}. In the next section we use $\hat{P}_{0,\omega}(\bm{r},\bm{r}')$ to calculate
the diffuson.

\subsection{Diffusion in disordered superconductors: The diffuson}\label{sec:diffuson}

The diffuson $\hat{P}_{d,\omega}(\bm{r},\bm{r}')$ is the classical probability of propagation from $\bm{r}$ to $\bm{r}'$ accounting for all paths including at least one scattering event. Summation over these paths is performed in the ladder approximation, as sketched in \autoref{fig:ladder}, giving the  equation
\begin{align}\label{eq:pddeffull}
&\hat{P}_{d,\omega}(\bm{r},\bm{r}')= \\ &\int\!d^{d}r_1\int\! d^{d}r_2 \hat{P}_{0,\omega}(\bm{r},\bm{r}_1)\hat{\Gamma}_{\omega}(\bm{r}_1,\bm{r}_2)\hat{P}_{0,\omega}(\bm{r}_2,\bm{r}'). \nonumber
\end{align}%
The Drude-Boltzmann factors account for the trajectory before the first scattering event and after the last one, at $\bm{r}_1$ and $\bm{r}_2$ respectively. The structure factor $\hat{\Gamma}_\omega(\bm{r}_1,\bm{r}_2)$ includes all scattering events. In our formalism, it is a 4$\times$4 matrix defined self-consistently by
\begin{align}\label{eq:gamdeffull}
&\hat{\Gamma}_\omega(\bm{r}_1,\bm{r}_2)= \\ & \gamma_e\hat{U}_v\big[\delta^{(d)}(\bm{r}_1-\bm{r}_2)+\int d^dr'' \hat{P}_{0,\omega}(\bm{r}_1,\bm{r}'')\hat{\Gamma}_\omega(\bm{r}'',\bm{r}_2) \big],\nonumber
\end{align}%
see the bottom half of  \autoref{fig:ladder}.
The Drude-Boltzmann contribution decays on a length scale of the order of the mean free path $l_e$, cf. Eq.~(\ref{eq:greenr}). Here we are interested in the diffusive regime, where the length scale $\lambda$ over which the structure factor varies is much longer than the mean free path, $\lambda \gg l_e$. We can then approximate $\hat{\Gamma}_\omega(\bm{r}_1,\bm{r}_2) \approx \hat{\Gamma}_\omega(\bm{r},\bm{r}')$. In this limit, \autoref{eq:pddeffull} can be approximately rewritten as
\begin{equation}
\hat{P}_{d,\omega}(\bm{r},\bm{r}')=\langle \hat{P}_{0} \rangle_{\bm{r}} \hat{\Gamma}_\omega(\bm{r},\bm{r}') \langle \hat{P}_{0} \rangle_{\bm{r}},
\label{eq:pddefp0gp0}
\end{equation}%
with $\langle \hat{P}_{0} \rangle_{\bm{r}}\equiv \langle \hat{P}_{0,\omega=0} \rangle_{\bm{r}}$ and
\begin{equation}
    \langle \hat{P}_{0,\omega} \rangle_{\bm{r}} \equiv \int d^dr' \hat{P}_{0,\omega}(\bm{r},\bm{r}').
    \label{eq:p0r}
\end{equation}%

Diffusion takes place at sufficiently long times beyond the scale $\tau_\text{min}$ so that terms of the order $(\omega\tau_\text{min})^2$ and $(\omega\tau_\text{min})(l_e/\lambda)^2$ can be neglected in comparison to those of order $(\omega\tau_\text{min})$ and $(l_e/\lambda)^2$, respectively. For the diffusion, in a normal metal the scale $\tau_\text{min}$ is simply given by $\tau_e= l_e/v_F$. Analogously, for a superconductor, we  obtain the scattering time \begin{equation}\label{eq:taus}
  \tau_s = \frac{l_e}{v_g}\, , \quad v_g = v_F \frac{\epsilon}{|E|}
\end{equation}
with $v_g$ the group velocity of the quasiparticles. However, we find that diffusion only sets in after the longer time
\begin{equation}\label{eq:taumindef}
    \tau_\text{min} = \max\left\{\tau_s,\frac{\Delta}{\epsilon^2}\right\}\, .
\end{equation}
The second scale $\Delta/\epsilon^2$ appears in order that the diffusive modes [first two entries of Eq.~\eqref{eq:mmat} below] are decoupled from the massive modes (last two entries). The definition in Eq.~(\ref{eq:taumindef}) reduces to $\tau_\text{min}=\tau_e$ in the normal state, while in the superconducting one we find
\begin{equation}
\tau_{\text{min}}=\begin{cases}
\frac{\Delta}{\epsilon^2},  & E<E_*,\\
\tau_s, & E>E_*.
     \end{cases}
    \label{eq:taumin}
\end{equation}%
where $E_*$ is defined as the energy at which $\tau_s=\Delta/\epsilon^2$. The magnitude of $E_*$ is sensitive to the disorder strength in the superconductor. Writing $E_*=\Delta + \varepsilon_*$, we obtain
\begin{equation}
\varepsilon_*\simeq \begin{cases}
\sqrt{\frac{\Delta}{\tau_e}}, & \tau_e \Delta \ll 1,\\
\frac{1}{2\Delta \tau_e^2} , & \tau_e \Delta \gg 1 .
     \end{cases}
    \label{eq:varepsilonstar}
\end{equation}%
where the condition $\tau_e \Delta \ll 1$ identifies the dirty regime, in which $\varepsilon_* \gg \Delta$, and $\tau_e \Delta \gg 1$ the clean case, where $\varepsilon_* \ll \Delta$. We will discuss in Sec.~\ref{sec:thermalconductivity} the relevance of this and other energy scales to the thermal conductivity.

To obtain the diffusion equation for the structure factor $\hat{\Gamma}_\omega$, we expand the latter up to second order in $\bm{r}''-\bm{r}_1$ around $\bm{r}''= \bm{r}_1$ in the left hand side of \autoref{eq:gamdeffull}. That equation can then be cast in the form
\begin{equation}
\hat{M}_\omega(\bm{r})\hat{\Gamma}_\omega(\bm{r},\bm{r}')=\gamma_e \delta^{(d)} (\bm{r}'-\bm{r}),
\label{eq:gameq44}
\end{equation}%
with the 
matrix operator
\begin{equation}\hat{M}_{\omega}(\bm{r})=\hat{U}_v^{-1}-\gamma_e\langle \hat{P}_{0,\omega} \rangle_{\bm{r}} -\frac{\gamma_e}{2 d }\langle r^2 \hat{P}_{0} \rangle_{\bm{r}} \nabla^2_{\bm{r}} .
\label{eq:mdef}
\end{equation}%
We have again neglected terms of order $(\omega\tau_\text{min})(l_e/\lambda)^2$ by evaluating $\langle r^2 \hat{P}_{0,\omega} \rangle_{\bm{r}} $ at $\omega = 0$, which we denote by removing the $\omega$ subscript. The integration over space of the Drude-Boltzmann contribution $\langle \hat{P}_{0,\omega}\rangle_{\bm{r}}$ can be performed directly using \autoref{eq:p0def} and the Green's function in real space [we remind that in the diffusive regime we only need to keep terms of order $(\omega\tau_\text{min})^0$ and $(\omega\tau_\text{min})^1$]. Using the relation $\langle r^2 \hat{P}_{0}\rangle_{\bm{r}} = 2l_e^2 \langle \hat{P}_{0} \rangle_{\bm{r}}$ and the definition of the potential matrix $\hat{U}_v$ [see the text after \autoref{eq:disorder}], the
matrix operator $\hat{M}_{\omega}(\bm{r})$ is obtained straightforwardly.

We wish to study the structure of $\hat{M}_{\omega}(\bm{r})$ to understand the diffusive modes of $\hat{\Gamma}_{\omega}(\bm{r},\bm{r}')$. It is convenient to introduce the states $\ket{a_\pm}=\frac{1}{\sqrt{2}}(\ket{\text{e},\text{e}}\pm \ket{\text{h},\text{h}})$ and $\ket{b_\pm}=\frac{1}{\sqrt{2}}(\ket{\text{e},\text{h}}\pm \ket{\text{h},\text{e}})$. 
We then work in the basis $B=\{\ket{a_-}, \text{cos}(\theta)\ket{a_+}+ \text{sin}(\theta)\ket{b_+}, \text{cos}(\theta)\ket{b_+}-\text{sin}(\theta)\ket{a_+},\ket{b_-}\}$, where $\Delta/E=\text{tan}(\theta)$.
In this basis, the structure of $\hat{M}_\omega(\bm{r})$ simplifies and the behavior of the diffusive modes can be singled out. 
Indeed, we find in the diffusive regime the result
\begin{equation}
\hat{M}_\omega(\bm{r})= \operatorname{diag}\left(\tau_s\mathcal{D}, \tau_s\tfrac{ \epsilon^2}{E^2+\Delta^2}\mathcal{D},-\tfrac{E^2+\Delta^2}{\epsilon^2},-1 \right),
\label{eq:mmat}\end{equation}%
with $\mathcal{D}= - i \omega-D_s\nabla^2_{\bm{r}}$. Here $D_s$ is the superconducting diffusion constant
\begin{equation}
D_s=\frac{v_g l_e}{d}
    \label{eq:ds}
\end{equation}
which, similarly to the scattering rate above, is energy dependent. On the other hand, the mean free path, proportional to $\sqrt{D_s \tau_s}$, remains independent of energy and equal to that in the normal state. These findings are in agreement with those in Ref.~\cite{BRT}.

Equation~(\ref{eq:mmat}) shows that in the diffusive limit $\hat{M}_\omega(\bm{r})$ is a diagonal matrix with two diffusive and two fast modes. We will  neglect the fast modes and focus on the diffusive ones. To this end, we define $\hat{\mathsf{M}}_{\omega}(\bm{r})$ as the 2$\times$2 matrix obtained by projecting $\hat{M}_\omega(\bm{r})$ into the 
subspace spanned by $\{\ket{a_-},\text{cos}(\theta)\ket{a_+}+\text{sin}(\theta)\ket{b_+}\}$.
According to \autoref{eq:gameq44}, the structure factor $\hat{\mathsf{\Gamma}}_\omega(\bm{r},\bm{r}')$ in this subspace satisfies the equation
\begin{equation}
\left[\tau_s
\begin{pmatrix}1 & 0 \\ 0 &  \frac{\epsilon^2}{E^2+\Delta^2}\end{pmatrix} \mathcal{D}\right]\hat{\mathsf{\Gamma}}_{\omega}(\bm{r},\bm{r}')=\gamma_e \delta^{(d)}(\bm{r}'-\bm{r}), \label{eq:gameq22}
\end{equation}%
where the terms in square brackets are the matrix $\hat{\mathsf{M}}_{\omega}(\bm{r})$.
We can rewrite \autoref{eq:pddefp0gp0} using $\hat{\mathsf{\Gamma}}_{\omega}(\bm{r},\bm{r}')$ as
\begin{equation}
\hat{P}_{d,\omega}(\bm{r},\bm{r}')=\hat{P}_v \hat{\mathsf{\Gamma}}_{\omega}(\bm{r},\bm{r}') \hat{P}_v^T,   \label{eq:pddefpvgph}
\end{equation}%
where $\hat{P}_v$ is the matrix with dimension 4$\times$2 that encompasses the first two columns of $ \langle \hat{P}_{0} \rangle_{\bm{r}}$ in the previously introduced basis $B$; it has the useful property $\gamma_e^2 \hat{P}_v^T \hat{P}_v =1$.
The diffuson thus found is a rank two matrix that takes the form $\hat{P}_{d,\omega}(\bm{r},\bm{r}')=\operatorname{diag}[\hat{P}_{d,\omega}(\bm{r},\bm{r}')_{1,1},\hat{P}_{d,\omega}(\bm{r},\bm{r}')_{2,2},0,0]$ in the basis $\tilde{B}=\{\ket{a_-}, \text{cos}(\theta)\ket{a_+}- \text{sin}(\theta)\ket{b_+}, \text{cos}(\theta)\ket{b_+}+\text{sin}(\theta)\ket{a_+},\ket{b_-}\}$, which is also the eigenbasis of $\langle \hat{P}_{0,\omega} \rangle_{\bm{r}}$. The 2$\times$2 upper left submatrix $\hat{\mathsf{P}}_{d,\omega}(\bm{r},\bm{r}')=\operatorname{diag}[\hat{P}_{d,\omega}(\bm{r},\bm{r}')_{1,1},\hat{P}_{d,\omega}(\bm{r},\bm{r}')_{2,2}]$  follows a diffusion equation given by
\begin{equation}
\frac{v_F}{2 \pi \rho_0 v_g}
\begin{pmatrix}1 & 0 \\ 0 & \frac{\epsilon^2}{E^2+\Delta^2}\end{pmatrix}\mathcal{D}\,\hat{\mathsf{P}}_{d,\omega}(\bm{r},\bm{r}')=\delta^{(d)}(\bm{r}'-\bm{r}).   \label{eq:pdeq22}
\end{equation}%
The result resembles the diffuson in the normal metal, but in the superconduncting state the diffusion constant and the scattering time depend on the group velocity $v_g$ which is no longer equal to the Fermi velocity [\autoref{eq:pdeq22} can also be reformulated to include the energy scaling $v_g/v_F$ in the frequency component rather than in the diffusion constant].
After applying the temporal Fourier transform, we obtain a direct relation between the probabilities of diffusion in the superconducting and normal states

\begin{equation}
\begin{split}
&\bra{i',j'}\hat{P}_d(\bm{r},\bm{r}';t)\ket{i,j}=P_{\text{n}}(\bm{r},\bm{r}';t v_g /v_F)\\&\frac{\pi \rho_0 v_g}{(E^2-\Delta^2)v_F}\big[(2E^2-\Delta^2)\delta_{i,j}\delta_{i,i'}\delta_{j,j'}+\\&\Delta^2(1-\delta_{i,i'})(1-\delta_{j,j'})+\Delta^2\delta_{i,i'}\delta_{j,j'}(1-\delta_{i,j})\\&-\Delta E (\delta_{i,i'}(1-\delta_{j,j'})+\delta_{j,j'}(1-\delta_{i,i'}))\big],
 \label{eq:pscpnm}
 \end{split}
\end{equation}
where $i,j,i',j' \in \{\text{e},\text{h}\}$ and the normal-state diffusion probability satisfies the equation
\begin{equation}
\left(-D\nabla^2_{\bm{r}}+\frac{\partial}{\partial t}\right)P_{\text{n}}(\bm{r},\bm{r}';t)=\delta^{(d)}(\bm{r}'-\bm{r})\delta(t).  \label{eq:pdiffnm}
\end{equation}%
where the diffusion constant $D$ coincides with the limit of zero order parameter for $D_s$ of Eq.~(\ref{eq:ds}) (in which case $v_g \to v_F$). We note  that the factor $v_F/2\pi \rho_0 v_g$ appearing in \autoref{eq:pdeq22} and \autoref{eq:pscpnm} is due to the unconventional normalization used for the probability. Since our main interest is the calculation of the thermal conductivity it is more convenient to directly calculate the disorder average product of Green's function which do not correspond to the normalized probability of diffusion. More details on the normalization are given in Appendix~\ref{appendix:particlediffusion}.
Equations (\ref{eq:gameq22}) and (\ref{eq:pdeq22}) can also be obtained in momentum space: by inverting matrix $\hat{M}_{\omega}(\bm{q})$, the calculation of both $\hat{\Gamma}_\omega(\bm{q})$ and $\hat{P}_{d,\omega}(\bm{q})$ is straightforward (see Appendix~\ref{appendix:kspace}).
Relations between diffusion in the superconducting and normal states similar to \autoref{eq:pscpnm} have been recently obtained for energies below the gap and at $\omega=0$~\cite{KBN}. We note that such subgap (virtual) diffusion can mediate the exchange interaction between two spin qubits tunnel-coupled to a superconductor~\cite{HCB}.

\subsection{Weak localization: The Cooperon}\label{sec:Cooperon}

\begin{figure}[tb]
  \includegraphics[width=0.85\linewidth]{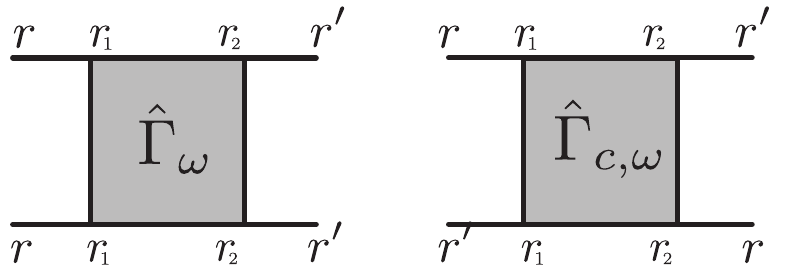}
    \caption{Diagrammatic representation of the diffuson (left) and the Cooperon (right). The Cooperon corresponds to reversing one of the trajectories that take part in the probability so that the impurities are encountered in inverse order. The upper Green's functions are retarded while the lower ones are advanced.}
    \label{fig:coop}
\end{figure}

After studying classical diffusion within the ladder approximation, we now focus on the first quantum correction to the probability of diffusion arising from localization effects (the Cooperon contribution).
The Cooperon matrix $\hat{P}_{c,\omega}(\bm{r},\bm{r}')$, shown schematically next to the diffuson in \autoref{fig:coop}, corresponds to the quantum interference between two trajectories covering the exact same path but in opposite directions. This interference effect is reflected in the structure of the expression
\begin{widetext}
\begin{eqnarray}
\hat{P}_{c,\omega}(\bm{r},\bm{r}')&=&\int d^dr_1 \int d^dr_2 \big(\overline{\hat{G}^R_{E +  \omega}(\bm{r},\bm{r}_1)}\otimes\overline{\hat{G}^A_E(\bm{r}',\bm{r}_1)^T}\big)\hat{\Gamma}_{c,\omega}(\bm{r}_1,\bm{r}_2)\big(\overline{\hat{G}^R_{E+ \omega}(\bm{r}_2,\bm{r}')}\otimes\overline{\hat{G}^A_{E}(\bm{r}_2,\bm{r})^T}\big),
\label{eq:pcdeffull}
\\
\hat{\Gamma}_{c,\omega}(\bm{r}_1,\bm{r}_2)&=&\gamma_e \hat{U}_v\big[\delta^{(d)}(\bm{r}_1-\bm{r}_2)+\int d^dr'' \big(\overline{\hat{G}^R_{E+ \omega} (\bm{r}_1,\bm{r}'')}\otimes \overline{\hat{G}^A_{E}(\bm{r}_1,\bm{r}'')^T}\big)\hat{\Gamma}_{c,\omega}(\bm{r}'',\bm{r}_2)\big].\label{eq:gamcdef}
\end{eqnarray}
\end{widetext}

Since the disorder-averaged Green's functions decay exponentially in real space [cf. Eq.~(\ref{eq:greenr})], it can already be seen above that the Cooperon is exponentially suppressed in $|\bm{r}'-\bm{r}|/l_e$.
We can simplify \autoref{eq:gamcdef} by noting that for a time reversal invariant system, \textit{i.e.}, for $\hat{G}^R_E(\bm{r},\bm{r}')=\hat{G}^R_E(\bm{r}',\bm{r})$, it is identical to \autoref{eq:gamdeffull} and thus $\hat{\Gamma}_{c,\omega}(\bm{r}_1,\bm{r}_2)=\hat{\Gamma}_\omega(\bm{r}_1,\bm{r}_2)$. We now again assume 
the latter to vary slowly on the scale of the mean free path and thus make the approximation $\hat{\Gamma}_{c,\omega}(\bm{r}_1,\bm{r}_2) \approx \hat{\Gamma}_\omega(\bm{r},\bm{r})$ in \autoref{eq:pcdeffull} which, neglecting terms of order $(\omega\tau_\text{min})^2$, $(\omega\tau_\text{min})(l_e/\lambda)^2$, and higher, becomes approximately
\begin{equation}
\hat{P}_{c,\omega}(\bm{r},\bm{r}')=\hat{F}(\bm{R})\hat{\Gamma}_{\omega}(\bm{r},\bm{r})\hat{F}(\bm{R}).
\label{eq:pcdeffgf}
\end{equation}%
Here, we define $\bm{R}=\bm{r}'-\bm{r}$ and
\begin{equation}
\begin{split}
\hat{F}(\bm{R})&=\int d^dr_1 \, \overline{\hat{G}_{E}^R(\bm{r},\bm{r}_1)}\otimes \overline{\hat{G}_E^A(\bm{r}',\bm{r}_1
)^T}\\&=\int \frac{d^dk}{(2\pi)^d}\, e^{i\bm{k}\cdot\bm{R}}\overline{\hat{G}_{E}^R(k)}\otimes \overline{\hat{G}_E^A(-k)^T},
\label{eq:Frdeffull}
\end{split}
\end{equation}%
which can be calculated by direct integration using Eq.~(\ref{eq:greenk}). Note the similarity between Eqs.~(\ref{eq:pddefp0gp0}) and (\ref{eq:pcdeffgf}), which become equivalent for $\bm{r}=\bm{r}'$, since $\langle \hat{P}_{0}\rangle_{\bm{r}} = \hat{F}(0)$. In fact,
there exists a general relation between the Cooperon and the diffuson of the form
\begin{equation}\hat{P}_{{c,\omega}}(\bm{r},\bm{r}')= \hat{P}_{{d,\omega}}(\bm{r},\bm{r}) \hat{f}(\bm{R}),
\label{eq:pcpdf}\end{equation}%
with $\hat{f}(\bm{R})$ given explicitly in Appendix~\ref{appendix:Cooperon}.

In the following, we focus on the element $\bra{a_-}\hat{P}_{c,\omega}(\bm{r},\bm{r}')\ket{a_-}$, which we denote as $\hat{P}_{{c,\omega}}(\bm{r},\bm{r}')_{1,1}$. This element is of particular interest since in the next Section it will be related to the thermal conductivity. We find%
\begin{equation}\hat{P}_{{c,\omega}}(\bm{r},\bm{r}')_{1,1}= \hat{P}_{{d,\omega}}(\bm{r},\bm{r})_{1,1} \hat{f}(\bm{R})_{1,1},
\label{eq:pc11pd11f11}\end{equation}%
with%
\begin{equation}
\hat{f}(\bm{R})_{1,1} \approx
     \begin{cases}
    \frac{1}{2}   e^{-\frac{R}{l_e}}  & (\text{1D}), \\
       \frac{1}{\pi k_F R }e^{-\frac{R}{l_e}}  & (\text{2D})\\
     \frac{1}{ 2 k_F^2 R^2 } e^{-\frac{R}{l_e}}, & (\text{3D}).\\
     \end{cases}
     \label{eq:fr}
\end{equation}%
where we have assumed $k_F R \gg 1$ and $\mu \gg \epsilon$,
and we have averaged fast oscillations over a spatial region of extension large compared to the Fermi wavelength $1/k_F$ but small compared to the mean free path $l_e$.
The weak localization correction \autoref{eq:pc11pd11f11} is a positive contribution to the probability of diffusion that is negligible when $R \gg l_e$. Consequentially, particles have an enhanced probability of returning to the origin. Due to conservation of the total probability, this implies a reduced probability of diffusion over long distances. This effect will be seen as a decrease of the thermal conductivity in \autoref{sec:thermalconductivity} and is qualitatively the same effect that the WL correction has on the transport coefficients of a normal metal.

The condition for the validity of the diffusive approximation $\omega\tau_\mathrm{min} \ll 1$
affects the return probability $\hat{P}_{d,\omega}(\bm{r},\bm{r})$. Based on that condition, the diffusive behavior of the system breaks down when considering very short timescales. On the other hand, on long time scales diffusion is limited by  the phase-coherence time $\tau_\phi$ (which can be related to the phase-coherence length via $L_\phi=\sqrt{D_s\tau_\phi}$).
The return probability at zero frequency is then given by%
\begin{equation}
    \hat{P}_d(\bm{r},\bm{r})=\int_{\tau_{\text{min}}}^{\tau_\phi}\! dt\,\hat{P}_d(\bm{r},\bm{r};t),
    \label{eq:pdtint}
\end{equation}%
where $\hat{P}_d(\bm{r},\bm{r}';t)$ is given in \autoref{eq:pscpnm}. Solving the diffusion equation (\ref{eq:pdiffnm}) in $d$-dimensional free space we obtain
\begin{equation}
    \hat{P}_d(\bm{r},\bm{r}';t)_{1,1}=\frac{2 \pi \rho_0 v_g}{v_F} \frac{1}{(4\pi D_s t)^{d/2}}e^{-R^2 /(4 D_s t)},
    \label{eq:pdt11}
\end{equation}%
where, as mentioned above, we 
focus for later use on the element $\hat{P}_{d}(\bm{r},\bm{r}';t)_{1,1}=\bra{a_-}\hat{P}_{d}(\bm{r},\bm{r}';t)\ket{a_-}$.
Inserting this result into \autoref{eq:pdtint} and performing the integral yields the return probability at zero frequency%
\begin{equation}
    \hat{P}_d(\bm{r},\bm{r})_{1,1}=\frac{4 \pi \rho_0}{D(4\pi)^{d/2}}D_s^{1-d/2}\begin{cases}
    \sqrt{\tau_{\phi}}-\sqrt{\tau_{\text{min}}}  & (\text{1D}), \\
      \ln\left(\frac{\sqrt{\tau_\phi}}{\sqrt{\tau_{\text{min}}}}\right)  & (\text{2D}),\\
    \frac{1}{\sqrt{\tau_{\text{min}}}}-\frac{1}{\sqrt{\tau_\phi}} & (\text{3D}).\\
     \end{cases}
    \label{eq:pdrrw0}
\end{equation}%
We remind that, unlike in the normal state where $\tau_\mathrm{min} = \tau_e$, in the superconducting state $\tau_{\text{min}}$ of Eq.~(\ref{eq:taumindef}) is an energy-dependent quantity. The dependence is qualitatively different in the two regimes separated by the energy $E_*$ [see Eq.~(\ref{eq:varepsilonstar})], and the energy $E_*$ itself takes different values in the clean and dirty regimes.

\section{\label{sec:thermalconductivity}Thermal conductivity}

In this section, we connect the results of the previous section concerning particle propagation to the thermal conductivity which is a physical observable. We obtain quantum corrections to the known results for the Drude-Boltzmann contribution \cite{Ambegaokar}. In particular, we derive explicit results for the weak localization correction to the thermal conductivity in two dimensions. Interestingly, in the superconducting state this correction displays a temperature dependence that differs from that in the normal state (or its simple extension to be discussed below).
Different regimes arise depending on the relations between temperature $T$, order parameter $\Delta(T)$, and the energy scale $\varepsilon_* = E_*-\Delta$.

\begin{figure}[tb]
  \includegraphics[width=\linewidth]{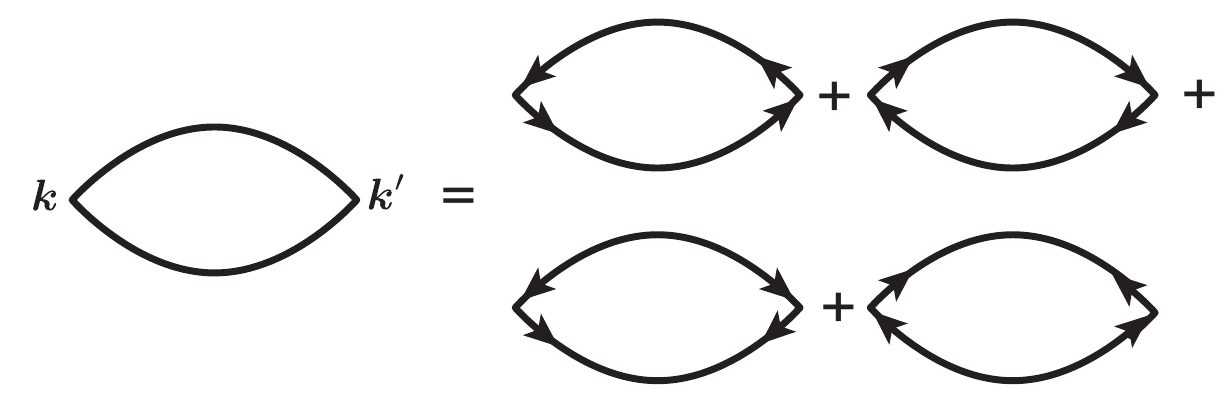}
    \caption{Diagrammatic representation of the thermal conductivity given by \autoref{eq:kdef} as the four specific components included in $\operatorname{Tr}[\overline{\tau_3 \operatorname{Im}\hat{G}^{R}_{E}(\bm{k},\bm{k}')\tau_3 \operatorname{Im}\hat{G}^{R}_{E}(\bm{k}',\bm{k})}]$.}
    \label{fig:feynman2}
\end{figure}

Our starting point is Kubo's formula for the thermal conductivity $K(T)$~\cite{Luttinger}; it can be written in terms of a product of Green's functions~\cite{Ambegaokar}
\begin{equation}
K=\frac{1}{4  \pi k_B T^2 m^2} \int_{\Delta}^\infty \! dE\,   \frac{E^2}{\cosh^2(\frac{E}{2k_BT})} I,
\label{eq:kdef}
\end{equation}%
with~\footnote{Note that the general expression for the Green's functions in momentum space depends on both the initial and final momentum $\bm{k}$ and $\bm{k}'$. The simplified expression given in \autoref{eq:greenk} assumes that after disorder averaging the Green's functions are $\propto\delta_{\bm{k},\bm{k}'}$}
\begin{equation}
I=\! \int \!\! \frac{d^d k}{(2\pi)^d}\frac{d^d k'}{(2\pi)^d}k_xk'_x \!\operatorname{Tr}[\overline{\tau_3 \! \operatorname{Im}\hat{G}^R_{E}(\bm{k},\bm{k}')\tau_3 \! \operatorname{Im}\hat{G}^R_E(\bm{k}',\bm{k}
)}],
\label{eq:Idef}
\end{equation}%
where $k_B$ is the Boltzmann constant and we take $x$ as the direction of the temperature gradient (and hence of heat propagation in an isotropic material, to which we restrict our attention). Its diagrammatic representation can be seen in \autoref{fig:feynman2}.

Using the above expression, we rewrite $I = I_A - I_D$ as the difference between two integrals with%
\begin{equation}
I_{A}= \text{Re} \! \int \! \frac{d^d k}{(2\pi)^d}\frac{d^dk'}{(2\pi)^d}\frac{k_xk'_x}{2}\operatorname{Tr}\!\big[\overline{\tau_3  \hat{G}^R_{E}(\bm{k},\bm{k}')\tau_3 \hat{G}^A_E(\bm{k}',\bm{k})}\big],
\label{eq:ia}
\end{equation}%
\begin{equation}
I_{D}=\text{Re}\! \int \! \frac{d^d k}{(2\pi)^d}\frac{d^dk'}{(2\pi)^d}\frac{k_xk'_x}{2}\operatorname{Tr}\!\big[\overline{\tau_3  \hat{G}^R_{E}(\bm{k},\bm{k}')\tau_3 \hat{G}^R_E(\bm{k}',\bm{k})}\big].
\label{eq:id}
\end{equation}%
In the regime $\mu\gg \epsilon,\,\Delta$ discussed after Eq.~(\ref{eq:dbar}), we can approximate $k_x \approx k_F u_x$, where $\bm{u}$ is the unit vector on the Fermi surface, and similarly for $k'_x$. Therefore, only the relative angle between the two momenta $\bm k, \bm k' $  matters.
Indeed, we discuss below the dependence of the disorder-averaged product of Green's functions on the relative orientation of $\bm{k}$ and $\bm{k}'$.
Once this dependence is known,
the integrals $I_A$ and $I_D$ can then be related to the so called $A$-type and $D$-type diffusive modes, respectively \cite{AZ}. The $A$-type modes contribution $I_A$ is proportional to the probability of diffusion $\hat{P}_{\omega}(\bm{r},\bm{r}')$ studied in \autoref{sec:disorder}, and the $D$-type modes one to $\hat{P}^D_{\omega}(\bm{r},\bm{r}')=\overline{\hat{G}^R_{E+\omega}(\bm{r},\bm{r}')\otimes \hat{G}_E^R(\bm{r}',\bm{r})}$. In systems with time reversal symmetry it can be shown that the $D$-type modes do not contribute to the thermal conductivity, \textit{i.e}, $I_D=0$ \cite{Ambegaokar}. With $I=I_A$ we can calculate the thermal conductivity using the results from the previous section.
We only briefly sketch  how to use those results here; more details on how to relate transport coefficients to the propagation probability can be found in Ref.~\cite{AM}.

As done for the total probability of diffusion, we divide the different contributions to the thermal conductivity into $K_0$, $K_d$ and $K_c$ and define as such the integrals $I_{0}$, $I_{d}$ and $I_{c}$. These integrals are related to the quantities $\hat{P}_{0}(\bm{r},\bm{r}')_{1,1}$, $\hat{P}_{d}(\bm{r},\bm{r}')_{1,1}$, and $\hat{P}_{c}(\bm{r},\bm{r}')_{1,1}$ defined in the previous section; here we use the identity $\hat{P}(\bm{r},\bm{r}')_{1,1}=\text{Tr}\big[\overline{\tau_3  \hat{G}^R_{E}(\bm{r},\bm{r}')\tau_3 \hat{G}^A_E(\bm{r}',\bm{r})}\big]/2$,
where we dropped the subscript $\omega=0$ to simplify the notation and we remind that the term on the left hand side is defined as $\hat{P}(\bm{r},\bm{r}')_{1,1}=\bra{a_-}\hat{P}(\bm{r},\bm{r}')\ket{a_-}$.
The Drude-Boltzmann integral $I_0$ represents propagation in a disordered medium without any scattering event taking place. In the absence of scattering, the initial and final momenta of the Green's functions are the same, with $\overline{\hat{G}^R_{E}(\bm{k},\bm{k}')}\otimes \overline{\hat{G}^A_E(\bm{k}',\bm{k})^T} \propto \delta^{(d)}(\bm{k}-\bm{k}')$. Then the angular integration in momentum space is equivalent to taking the product of momenta out of the integral as $k_F^2/d$, and the  relation between $I_0$ and $\hat{P}_0(\bm{r},\bm{r}')_{1,1}$ can be obtained by going into real space, using the Fourier transform for a translational invariant system
\begin{equation}
\int\frac{d^dk}{(2\pi)^d} \overline{\hat{G}^R_{E}(k)}\otimes \overline{\hat{G}^A_E(k)^T}=\int d^d r' \hat{P}_0(\bm{r},\bm{r}').
\label{eq:p0fourier}
\end{equation}
The Drude-Boltzmann integral is then
\begin{equation}
I_{0}=\frac{k_F^2}{d}\text{Re}\int d^dr'\hat{P}_{{0}}(\bm{r},\bm{r}')_{1,1},
\label{eq:i0}
\end{equation}%
and using Eq.~(\ref{eq:p0r}) [see also \autoref{eq:p0btilde}] we obtain $I=k_F^2/d \gamma_e$. Inserting the result into \autoref{eq:kdef} we obtain the Drude-Boltzmann contribution to the thermal conductivity
\begin{equation}
K_0=\frac{D\rho_0 }{2 k_B T^2}\int_\Delta^\infty dE \frac{E^2}{\text{cosh}^2\left(\frac{ E}{2k_BT}\right)},
\label{eq:k0}
\end{equation}%
where $D=v_F l_e/d$ is the diffusion constant in the normal state. This formula agrees with previous calculations~\cite{Ambegaokar}. It is equivalent to the result in the normal state with the sole difference that only states with energy $E> \Delta$ contribute. The absence of states below the gap is reflected in the lower limit of the integral and leads to the exponential suppression of $K_0$ at temperatures $k_BT\ll\Delta$.

For the diffuson integral $I_d$, we find simply $I_d=0$. This result is valid for isotropic scattering by impurities: the initial and final momenta of the Green's functions ($\bm{k}$ and $\bm{k}'$, respectively) have uncorrelated directions after a large number of scattering events, which leads to the vanishing of the angular integration in \autoref{eq:ia}. Anisotropic scattering would result in the substitution of the scattering time $\tau_e$ with the transport time in Eq.~(\ref{eq:k0})~\cite{AM} ($\tau_e$ enters that equation via the mean free path in the diffusion constant).

Similar considerations to those above make it possible to relate $I_c$ to $\hat{P}_{c}(\bm{r},\bm{r}')_{1,1}$. The Cooperon accounts for an enhanced probability of a particle to return to its initial point; therefore, its initial and final momenta will be approximately opposite to each other. The integrand of $I_c$ is then sharply peaked around $\bm{k}=-\bm{k}'$, and can be approximated to be proportional to $\delta^{(d)}(\bm{k}+\bm{k}')$~\footnote{This result can be obtained mathematically by calculating the structure factor for the Cooperon in momentum space. It has a peak at $\bm{k}+\bm{k}'=0$, with $\hat{\mathsf{\Gamma}}_{c,\omega}(\bm{k}+\bm{k}')=\hat{\mathsf{\Gamma}}_\omega(\bm{k}+\bm{k}')$, where $\hat{\mathsf{\Gamma}}_\omega(\bm{q})$ is defined in \autoref{eq:gamqmat}; see also Ref.~\cite{AM}.}. We again take the product of momenta out of the integral, and going over to real space yields
\begin{equation}
I_{c}=-\frac{k_F^2}{d}\text{Re}\int d^dr'\hat{P}_{{c}}(\bm{r},\bm{r}')_{1,1}.
\label{eq:icpc}
\end{equation}%
After substituting \autoref{eq:pc11pd11f11} into the above expression we have
\begin{equation}
I_{c}=-\frac{k_F^2}{d}\text{Re}\,\hat{P}_{{d}}(\bm{r},\bm{r})_{1,1} \int\!d^dR \, \hat{f}(\bm{R})_{1,1}.
\label{eq:icpdf}
\end{equation}%
Using the expressions for $\hat{f}(\bm{R})_{1,1}$ given in \autoref{eq:fr}, we find that for all dimensions
\begin{equation}
    \int\!d^dR\,\hat{f}(\bm{R})_{1,1}=\frac{\tau_e}{  \pi \rho_0},
    \label{eq:frrint}
\end{equation}%
and inserting these results into \autoref{eq:kdef}, we arrive at
\begin{equation}
K_c=-\frac{D}{4\pi^2 k_B T^2  \rho_0}\int_\Delta^\infty dE \frac{E^2}{\text{cosh}^2\left(\frac{ E}{2k_BT}\right)}\hat{P}_{{d}}(\bm{r},\bm{r})_{1,1},
\label{eq:kc}
\end{equation}%
where the return probability $\hat{P}_{{d}}(\bm{r},\bm{r})_{1,1}$, given by \autoref{eq:pdrrw0}, is a function of energy $E$. 
This energy dependence leads to a temperature dependence of $K_c$ which we study in the following for two dimensions. We note that it is crucial to retain this energy dependence. Neglecting the energy dependence of the return probability, we would find the incorrect result
$K_c/K_0 = - \hat{P}_{{d}}(\bm{r},\bm{r})_{1,1}/2\pi^2\rho_0^2$ 
and the temperature dependence of $K_c$ would simply follow from the one in the normal state.

\subsection{Regimes for the WL correction to the thermal conductivity}
\label{sec:regimes}

As remarked above, the dependence of the return probability $\hat{P}_{{d}}(\bm{r},\bm{r})_{1,1}$ on energy makes it possible for the WL correction $K_c$ to the thermal conductivity to have a temperature dependence that differs from that of the main (Drude-Boltzmann) contribution $K_0$. Here we explore when such a deviation takes place and under which conditions it could be observable. To this aim, let us introduce the temperature $T_\Delta$ defined by $k_B T_\Delta = \Delta(T_\Delta)$; for our purposes, the temperature dependence of the gap on temperature is approximately captured by the interpolation formula~\cite{Gross1986}
\begin{equation}
    \Delta(T) \approx 1.76k_B T_c \tanh \left(1.74 \sqrt{\frac{T_c}{T}-1}\right),
    \label{eq:deltat}
\end{equation}
with $T_c$ the critical temperature of the superconductor. From this expression we find $T_\Delta \approx 0.9 T_c$. Clearly, both $K_0$ and $K_c$ are exponentially suppressed in the low-temperature regime $T \ll T_\Delta$, see Eqs.~(\ref{eq:k0}) and (\ref{eq:kc}), making their accurate measurement challenging. Therefore, the high-temperature regime $T_\Delta \lesssim T < T_c$ is most relevant in order to observe the effects of weak-localization. For completeness, we consider both regimes below (details of the calculations are presented in Appendix~\ref{appendix:wl}).

A second relevant temperature scale, denoted by $T_*$, can be defined via the equation $k_B T_* = \varepsilon_*(T_*)$, where $\varepsilon_*$ depends on temperature through the gap $\Delta(T)$, see Eq.~(\ref{eq:varepsilonstar}). For dirty superconductors, $\tau_e \Delta(0) \ll 1$, we have $T_* \simeq T_c$, while for clean ones, $\tau_e \Delta(0) \gg 1$, we find $T_* \ll T_c$, indicating that qualitatively different behaviors can be expected in the two cases. Finally, with regard to the effect of phase coherence on $K_c$, we consider two possibilities, namely an energy independent coherence time $\tau_\phi$ or an energy independent coherence length $L_\phi = \sqrt{D_S \tau_\phi}$.  These two possibilities are equivalent in the normal state, but in the superconducting one they are not, due to the energy dependence of the diffusion constant $D_S$, Eq.~(\ref{eq:ds}).

\subsubsection{High-temperature regime}

In the high-temperature regime $T \gtrsim T_{\Delta}$, in order to find the leading contributions to the heat conductivity, we approximate $k_BT \gg \Delta$. Then the WL correction in this regime does not depend on the gap $\Delta$. Moreover, for a superconductor in the clean limit, since $\varepsilon_* \ll \Delta$, the relative correction coincides with the one in the normal state~\footnote{we note that sufficiently close to $T_c$, a cross-over from clean to dirty takes place when $\Delta\tau_e \sim 1$. However, the condition $\varepsilon_* \ll T$ is met, which ensures the validity of our results (see also the discussion for the dirty case).},
\begin{equation}
    \frac{K_c}{K_0}=\frac{-1}{\pi k_F l_e}\begin{cases}
     \ln\left(\frac{\tau_\phi }{\tau_e}\right)& \text{($\tau_\phi$ fixed)}, \\
     2\ln\left(\frac{L_\phi}{l_e}\right) &\text{($L_\phi$ fixed)}.
    \end{cases}
    \label{eq:nmratio}
\end{equation}%

The same expressions hold for a dirty superconductor sufficiently close to $T_c$, so that $\varepsilon_* \ll k_B T$, but since, as mentioned above,  $T_* \simeq T_c$, this result has very limited applicability. More interestingly, there exists an intermediate regime, $T_\Delta \lesssim T \lesssim T_*$, in which the WL correction depends on the ratio $k_BT/\varepsilon_*$,
\begin{equation}
    \frac{K_c}{K_0}=\frac{-1}{\pi k_F l_e}\begin{cases}
     \ln\left(\frac{\tau_\phi }{\tau_e}\right)+2\ln\left(\frac{k_BT}{\varepsilon_*}\right)& \text{($\tau_\phi$ fixed)}, \\
     2\ln\left(\frac{L_\phi}{l_e}\right)+2\ln\left(\frac{k_BT}{\varepsilon_*}\right)  & \text{($L_\phi$ fixed)}.
    \end{cases}
    \label{eq:dirtyht}
\end{equation}%
Note that, in the high-temperature regime, the temperature dependence of the WL correction  is insensitive to the assumption of energy-independent dephasing time vs. length; this can be traced back to the fact that at the relevant energy scale (given by temperature), we have for the group velocity $v_g \approx v_F$, see Eq.~(\ref{eq:taus}).

\subsubsection{Low-temperature regime}

In the low temperature regime $T \ll T_{\Delta}$ we have $k_BT \ll \Delta$, which results in the exponential suppression of both $K_0$ and $K_c$ discussed above. Their ratio, however, is not exponentially suppressed. Indeed,
the WL correction for a dirty superconductor is given by
\begin{equation}
    \frac{K_c}{K_0}=\frac{-1}{\pi k_F l_e}\begin{cases}
     \ln(\frac{\tau_\phi }{\tau_e})+\ln(\frac{\Delta k_BT}{\varepsilon_*^2})  & \text{($\tau_\phi$ fixed)}, \\
     2\ln(\frac{L_\phi}{l_e}) + \ln(\frac{\Delta^{3/2} \sqrt{k_BT}}{\varepsilon_*^2})  & \text{($L_\phi$ fixed)}.
    \end{cases}
    \label{eq:dirtylt}
\end{equation}%
In both cases, at the cross-over temperature $T_\Delta$ the correction agrees with that found in the high-temperature regime. However, the temperature dependence is now sensitive to the assumption of energy-independent dephasing time/length.

 For a clean superconductor in the regime $T_* < T < T_{\Delta}$, the normalized WL correction is
\begin{equation}
    \frac{K_c}{K_0}=\frac{-1}{\pi k_F l_e}\begin{cases}
     \ln(\frac{\tau_\phi}{\tau_e})+\frac12\ln(\frac{k_BT}{\Delta})  & \text{($\tau_\phi$ fixed)}, \\
     2\ln(\frac{L_\phi}{l_e})   & \text{($L_\phi$ fixed)}.
    \end{cases}
    \label{eq:cleanlt}
\end{equation}%
We note that, according to Eqs.~(\ref{eq:nmratio}) and (\ref{eq:cleanlt}), for $T>T_*$ and assuming energy-independent dephasing length, the WL correction in the clean case coincides with that in the normal state. This finding resembles that for the WL correction to the heat conductance of superconductor/normal/superconductor junctions with short (shorter than dephasing length) normal part in the absence of phase gradient and gap differences~\cite{HHS}. In that case, the latter two assumptions ensure that the transmission probability of quasiparticles excitations through the junction is independent of energy. Similarly here, the assumptions of energy-independent dephasing length and sufficiently high temperature ensure that the return probability of Eq.~(\ref{eq:pdrrw0}) is energy-independent over the relevant energy range.
For a clean superconductor there exists also a regime where $k_BT \ll \varepsilon_*$, where this energy independence does not hold.  This regime is calculated in Appendix~\ref{appendix:wl}, but we do not discuss it here further as it has a limited validity at temperatures where the thermal conductivity is strongly supressed~\footnote{We stress that all the results of this section are valid only under certain conditions on $\tau_\phi$ or $L_\phi$, explained in Appendix \ref{appendix:wl}, which ensure that the sum of the logarithms is positive.}.

\section{\label{sec:conclusiones}Summary and discussion}

\begin{figure}[tb]
  \includegraphics[width=\linewidth]{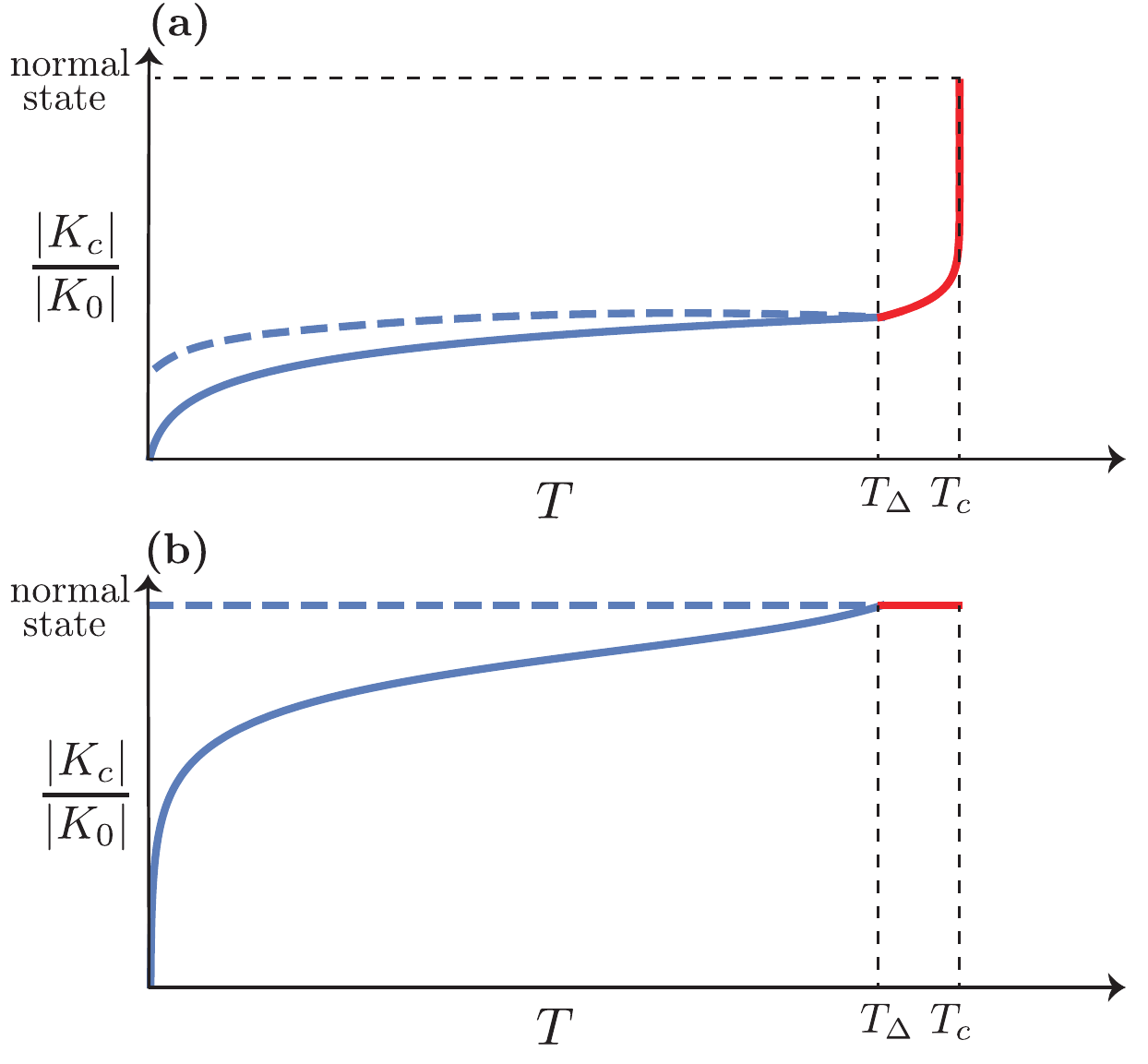}
    \caption{Schematic representation of the normalized weak localization correction $|K_c/K_0|$ as a function of temperature for an $s$-wave superconductor in the \textbf{(a)} dirty limit, where $T_* \approx T_c$ and \textbf{(b)} clean limit, where $T_* \ll T_c$. The solid lines represent the results for energy-independent dephasing time (fixed $\tau_\phi$) and the dashed lines for energy-independent dephasing length ($L_\phi$ fixed). The blue color highlights the behaviour in the low temperature regime $T<T_\Delta$, and the red color in the high temperature one.}
    \label{fig:temp}
\end{figure}%

In this work, we have calculated the weak localization correction to the thermal conductivity in conventional disordered superconductors. As our starting point, we have studied diffusion with the help of a general formalism based on semiclassical Green's functions and their corresponding matrix expressions in Nambu space, see Sec.~\ref{sec:disorder}. The formalism can be straightforwardly expanded to tackle systems with different symmetries; as an example, in Appendix~\ref{appendix:so} we investigate diffusion in the presence of weak spin-orbit scattering.

The thermal conductivity $K$ can be obtained from the probability of diffusion and, similarly to the calculation of electrical conductivity in the normal state, the weak localization correction can be related to the Cooperon $\hat{P}_{c}(\bm{r},\bm{r}')$, see \autoref{sec:Cooperon}. In fact, the correction always reduces the thermal conductivity which is consistent with the results for electrical conductivity in normal metals~\cite{AM,AltshulerAronov}. Our calculations in \autoref{sec:disorder} show that diffusion is reduced as the probability of return to the origin is increased due to WL.

As the temperature decreases below the critical temperature, the thermal conductivity is suppressed due to the opening of the gap $\Delta$ in the density of states; this leads to the well-known exponential suppression of $K$ at temperatures $T\ll T_\Delta \approx 0.9 T_c$. Interestingly, we find that the WL correction is affected not only by the gap, but also by a second energy scale $\varepsilon_*$ related to both the gap and the impurity scattering time $\tau_e$, see Eq.~(\ref{eq:varepsilonstar}). This energy scale encodes the fact that the onset of diffusion takes longer and longer times as the energy approaches the gap (while being limited only by the scattering time in the normal state); similarly, the diffusion constant decreases as energy decreases toward the gap [Eq.~(\ref{eq:ds})]. As a consequence, by lowering temperature the probability of return to the origin is decreased compared to the normal state, and the magnitude of the WL correction decreases.

For both clean ($\tau_e \Delta \gg 1$) and dirty  ($\tau_e \Delta \ll 1$) superconductors, we have considered the high ($T>T_\Delta$) and low ($T<T_\Delta$) temperature regimes, as summarized in Fig.~\ref{fig:temp} for two dimensions.
We highlight the regime $ T_\Delta < T < T_*\approx T_c$, which exists only in dirty superconductors, as the most interesting for the experimental verification of our results.
In this temperature range, the  thermal conductivity is not yet exponentially suppressed but, at the same time, most of the decrease in the magnitude of the WL correction has taken place, see Fig.~\ref{fig:temp}(a) and Eq.~(\ref{eq:dirtyht}). An interesting question for future research is the generalization of the approach presented here to calculate transport properties in disordered $d$-wave superconductors~\cite{GYSR,DL,YXLY}, for which the weak localization correction to thermal conductivity has so far been considered only in the mixed state~\cite{BCSZ}. For both $s$- and $d$-wave superconductors, calculating the effect of Zeeman splitting on the WL correction could also afford another avenue to experimentally check our theory.

\begin{acknowledgments}

This work was supported by the Deutsche Forschungsgemeinschaft (DFG) under Grant No. CA 1690/1.

\end{acknowledgments}

\appendix

\section{\label{appendix:particlediffusion}Normalization of the diffusion probability}

To discuss the normalization of the probability we consider particle conservation: in a superconducting system, the number of electrons plus the number of holes must be conserved. Let us define the two probabilities
\begin{eqnarray}
P_{e,\omega}(\bm{r},\bm{r}') & = & \overline{G^R_{E+ \omega}(\bm{r},\bm{r}')G^A_E(\bm{r}',\bm{r})}
\nonumber \\
& - & \overline{F^R_{E+ \omega}(\bm{r},\bm{r}')\bar{F}^{ A}_E(\bm{r}',\bm{r})},
\label{eq:pedef} \\
P_{h,\omega}(\bm{r},\bm{r}')& = & \overline{\bar{G}^{ R}_{E+ \omega}(\bm{r},\bm{r}')\bar{G}^{A}_E(\bm{r}',\bm{r}})
\nonumber \\
& - & \overline{\bar{F}^{R}_{E+ \omega}(\bm{r},\bm{r}')F^A_E(\bm{r}',\bm{r})}.
\label{eq:phdef}
\end{eqnarray}%
Here, $P_{e,\omega}(\bm{r},\bm{r}')$ is the probability that an electron propagates from $\bm{r}$ to $\bm{r}'$ plus the probability that said electron converts into a hole at some point during the trajectory. $P_{h,\omega}(\bm{r},\bm{r}')$ is the equivalent for holes.
These two quantities are related to $\hat{P}_{\omega}(\bm{r},\bm{r}')$ by
\begin{equation}
\bra{a_-}\hat{P}_{\omega}(\bm{r},\bm{r}')\ket{a_-}=\frac{1}{2}[P_{e,\omega}(\bm{r},\bm{r}')+P_{h,\omega}(\bm{r},\bm{r}')].
    \label{eq:pwpeph}
\end{equation}%
We can define the normalized probabilities $\mathcal{P}_{e,\omega}(\bm{r},\bm{r}')=A P_{e,\omega}(\bm{r},\bm{r}')$ and $\mathcal{P}_{h,\omega}(\bm{r},\bm{r}')=A P_{h,\omega}(\bm{r},\bm{r}')$ such that
\begin{equation}
\langle \mathcal{P}_{e,\omega} \rangle_{\bm{r}}=\langle \mathcal{P}_{h,\omega} \rangle_{\bm{r}}=\frac{i}{\omega},
\label{eq:normw}
\end{equation}%
which is the Fourier transform into frequency space of the normalization condition
\begin{equation}
\int d^dr' \mathcal{P}_{e,\omega}(\bm{r},\bm{r}';t)=\int d^dr' \mathcal{P}_{h,\omega}(\bm{r},\bm{r}';t)=1.
\label{eq:normt}
\end{equation}

Let us now consider the diffusion equation followed by $\hat{P}_{d,\omega}(\bm{r},\bm{r}')_{1,1}\equiv\bra{a_-}\hat{P}_{d,\omega}(\bm{r},\bm{r}')\ket{a_-}$, given by the first element of \autoref{eq:pdeq22}:
\begin{equation}
   \frac{v_F}{2\pi\rho_0 v_g}\left(
   - D_s \nabla^2_{\bm{r}}-i\omega\right)\hat{P}_{d,\omega}(\bm{r},\bm{r}')_{1,1}=\delta^{(d)}(\bm{r}'-\bm{r}).
   \label{eq:pd11eq}
\end{equation}%
After spatial integration, we find the normalization factor $A=v_F/2\pi \rho_0 v_g$.

\section{\label{appendix:kspace}Superconducting diffuson in momentum space}

To work in momentum space, we start by taking the Fourier transform \autoref{eq:gameq44}. The Laplace operator $\nabla^2_{\bm{r}}$ becomes the relative momentum squared $\bm{q}^2$, and $\hat{M}_{\omega}(\bm{q})$ can be inverted to obtain %
\begin{equation}
\hat{\Gamma}_\omega(\bm{q})=\gamma_e \hat{M}_{\omega}(\bm{q})^{-1}.
    \label{eq:gamqdef}
\end{equation}%
After calculating the inverse of $\hat{M}_{\omega}(\bm{q})$ explicitly, we can simplify it in the diffusive regime discussed in Sec.~\ref{sec:diffuson}, and $\hat{\Gamma}_\omega(\bm{q})$ is reduced to a rank two matrix whose non-zero elements correspond to $\hat{\mathsf{\Gamma}}_\omega(\bm{q})=\gamma_e \hat{\mathsf{M}}_{\omega}(\bm{q})^{-1}$, given in the basis $B_2=\{ \ket{a_-}, \text{cos}(\theta)\ket{a_+}+ \text{sin}(\theta)\ket{b_+} \}$ by%
\begin{equation}
\hat{\mathsf{\Gamma}}_\omega(\bm{q})=\frac{ \gamma_e}{\tau_s} \begin{pmatrix} \frac{1}{ D_sq^2 - i\omega} & 0 \\ 0 & \frac{E^2+\Delta^2}{\epsilon^2}\frac{1}{ D_sq^2 - i\omega}  \end{pmatrix}.
    \label{eq:gamqmat}
\end{equation}%
The diffuson, given by%
\begin{equation}\hat{P}_{d,\omega}(\bm{q})=\hat{P}_{0,\omega}(\bm{q})\hat\Gamma_\omega(\bm{q})\hat{P}_{0,\omega}(\bm{q}),\label{eq:pdqpqgpq}\end{equation}%
can be approximated in the limit of small relative momentum $\bm{q}$ and relative frequency $\omega$ as
\begin{equation}\hat{P}_{d,\omega}(\bm{q})=\hat{P}_0(0)\hat\Gamma_\omega(\bm{q})\hat{P}_0(0),
\label{eq:pdqp0gp0}
\end{equation}%
where $\hat{P}_0(\bm{q}=0)=\langle \hat{P}_0 \rangle_{\bm{r}}$, given in the original Nambu basis [defined after Eq.~(\ref{eq:greenra})] by
\begin{equation}\begin{split}
    \langle \hat{P}_0 \rangle_{\bm{r}}=&\frac{1}{2\gamma_e\epsilon^2} \times \\ &\begin{pmatrix}
    2E^2-\Delta^2 & -\Delta E & -\Delta E & \Delta^2 \\
    -\Delta E & \Delta^2 & \Delta^2 & -\Delta E \\
    -\Delta E & \Delta^2 & \Delta^2 & -\Delta E \\
    \Delta^2 & -\Delta E & -\Delta E & 2E^2-\Delta^2
    \end{pmatrix},
\end{split}
\label{eq:p0nambu}
\end{equation}
and in its eigenbasis $\tilde{B}$ by
\begin{equation}
    \langle \hat{P}_0 \rangle_{\bm{r}}=    \frac{1}{\gamma_e}\begin{pmatrix}
    1 & 0 & 0 & 0 \\
    0 & \frac{E^2+\Delta^2}{\epsilon^2} & 0 &  0 \\
    0 & 0 & 0 & 0 \\
    0 & 0 & 0 & 0
    \end{pmatrix}.
    \label{eq:p0btilde}
\end{equation}%
The diffuson can then be calculated by direct matrix multiplication. We obtain
\begin{equation}
\hat{P}_d(\bm{q})= \frac{1}{\tau_s}\frac{1}
{ D_sq^2-i\omega} \langle \hat{P}_0 \rangle_{\bm{r}},
 \label{eq:pdqp0r}
 \end{equation}
 which corresponds to a rank two matrix that can be written as
 \begin{equation}
\hat{\mathsf{P}}_{d,\omega}(\bm{q})=\frac{2\pi \rho_0 v_g}{v_F} \begin{pmatrix} \frac{1}{ D_sq^2 - i\omega} & 0 \\ 0 & \frac{E^2+\Delta^2}{\epsilon^2}\frac{1}{ D_sq^2 - i\omega}  \end{pmatrix}.
    \label{eq:pdqmat}
\end{equation}%
 in the basis $\tilde{B}_2=\{\ket{a_-}, \text{cos}(\theta)\ket{a_+}- \text{sin}(\theta)\ket{b_+})\}$. This is again equivalent to the result obtained by solving \autoref{eq:pdeq22} after performing a Fourier transform into momentum space.

\section{\label{appendix:Cooperon}Superconducting Cooperon}

Here we work out explicitly the relation between Cooperon and diffusion in the superconducting state. Since we are interested in the diffusive regime, the matrix $\hat{\Gamma}_{c,\omega}(\bm{r},\bm{r})=\hat{\Gamma}_\omega(\bm{r},\bm{r})$ can be simplified as a 2$\times$2 matrix that follows \autoref{eq:gameq22} in the subspace spanned by $B_2=\{\ket{a_-},\text{cos}(\theta)\ket{a_+}+\text{sin}(\theta)\ket{b_+}\}$. As done in Eq.~(\ref{eq:pddefpvgph}), we rewrite \autoref{eq:pcdeffgf} as%
\begin{equation}
    \hat{P}_{c,\omega}(\bm{r},\bm{r}')= \hat{F}_v(\bm{R}) \hat{\mathsf{\Gamma}}_{\omega}(\bm{r},\bm{r}) \hat{F}_v^T(\bm{R}),
    \label{eq:pcdeffvgfh}
\end{equation}%
where $\hat{F}_v(\bm{R})$ is defined, similarly to $\hat{P}_v$, as the matrix containing the first two columns of $\hat{F}(\bm{R})$ in the $B$ basis. By substituting the expression for $\hat{\mathsf{\Gamma}}_{\omega}(\bm{r},\bm{r})$ as a function of $\hat{P}_{d,\omega}(\bm{r},\bm{r})$ obtained from \autoref{eq:pddefpvgph}, we find%
\begin{equation}
    \hat{P}_{c,\omega}(\bm{r},\bm{r}')=\hat{A}(\bm{R})\hat{P}_{d,\omega}(\bm{r},\bm{r})\hat{A}(\bm{R})^T,
    \label{eq:pcdeffvphpdpvfh}
\end{equation}%
where%
\begin{equation}
\hat{A}(\bm{R})=\gamma_e^2 \hat{F}_v(\bm{R})\hat{P}_v^T.
    \label{eq:alpha}
\end{equation}%
The matrix $\hat{A}(\bm{R})$ is, like $\hat{P}_{d,\omega}(\bm{r},\bm{r})$, a rank two matrix whose only non-zero terms exist in the subspace spanned by the basis $\tilde{B}_{2}=\{\ket{a_-}, \text{cos}(\theta)\ket{a_+}- \text{sin}(\theta)\ket{b_+})\}$. The Cooperon $\hat{P}_{c,\omega}(\bm{r},\bm{r}')$ will therefore also share this property, and we can work with \autoref{eq:pcdeffvphpdpvfh} in the $\tilde{B}_{2}$ basis subspace to ensure the invertibility of all terms involved and simplify the calculation. We write this as
\begin{equation}
    \hat{\mathsf{P}}_{c,\omega}(\bm{r},\bm{r}')=\hat{\mathsf{A}}(\bm{R})\hat{\mathsf{P}}_{d,\omega}(\bm{r},\bm{r})\hat{\mathsf{A}}(\bm{R})^T,
    \label{eq:pcdeffvphpdpvfh22}
\end{equation}
where use of the sans serif fonts denotes the projection into the 2$\times$2 subspace.
We deduce from \autoref{eq:pdeq22} that $\hat{\mathsf{P}}_{d,\omega}(\bm{r},\bm{r})$ is diagonal and proportional to the matrix $\operatorname{diag}[1,\epsilon^2/(E^2+\Delta^2)]$ in the $\tilde{B}_{2}$ basis. We can then write in this basis
\begin{equation}
    \hat{\mathsf{P}}_{c,\omega}(\bm{r},\bm{r}')=\hat{\mathsf{P}}_{d,\omega}(\bm{r},\bm{r}) \hat{\mathsf{f}}(\bm{R})
    \label{eq:pcpdfappendix22}
\end{equation}%
where
\begin{equation}
    \hat{\mathsf{f}}(\bm{R})=\begin{pmatrix} 1 & 0 \\ 0 & \frac{E^2+\Delta^2}{\epsilon^2}\end{pmatrix}\hat{\mathsf{A}}(\bm{R})\begin{pmatrix} 1 & 0 \\ 0 & \frac{\epsilon^2}{E^2+\Delta^2}\end{pmatrix}\hat{\mathsf{A}}(\bm{R})^T.
    \label{eq:frmatrixdefinition}
\end{equation}
The equation in the full 4$\times$4 space can be obtained by expanding every matrix into the full $\tilde{B}$ basis by filling in zeroes in all the other elements of the matrix to obtain
\begin{equation}
    \hat{P}_{c,\omega}(\bm{r},\bm{r}')=\hat{P}_{d,\omega}(\bm{r},\bm{r}) \hat{f}(\bm{R}).
    \label{eq:pcpdfappendix}
\end{equation}%
In the two dimensional case, we have, in the $\tilde{B}_{2}$ basis
\begin{widetext}
\begin{equation}
    \hat{\mathsf{f}}(\bm{R})=e^{-R/l_e}\frac{1}{\pi k_F R}\begin{pmatrix}
   \text{cos}^2(k_h R-\frac{\pi}{4})+ \text{cos}^2(k_e R-\frac{\pi}{4}) &&   \frac{\sqrt{E^2+\Delta^2}}{\epsilon}\big[\text{cos}^2(k_h R-\frac{\pi}{4})- \text{cos}^2(k_e R-\frac{\pi}{4})\big] \\
    \frac{\epsilon}{\sqrt{E^2+\Delta^2}}\big[\text{cos}^2(k_h R-\frac{\pi}{4})- \text{cos}^2(k_e R-\frac{\pi}{4})\big] &&  \text{cos}^2(k_h R-\frac{\pi}{4})+ \text{cos}^2(k_e R-\frac{\pi}{4}) \end{pmatrix},
    \label{eq:frmat}
\end{equation}
\end{widetext}
where $k_e=k_F+\epsilon/ v_F$ and $k_h
=k_F-\epsilon/ v_F$.
We note that in contrast to
$\hat{P}_{d,\omega}(\bm{r},\bm{r}')$, this matrix is not diagonal in the $\tilde{B}$ basis;
that is, we have not fully separated the two low-energy modes.
However, we work in the limit $ \mu \gg \epsilon $, where the small difference in the frequency of oscillation between electrons and holes is negligible. The fast oscillations average out when integrating over
a length long compared to the Fermi wavelength but small compared to the mean free path, so that we can obtain an approximate formula for $\hat{\mathsf{f}}(\bm{R})$  by replacing
$\text{cos}^2(k_eR-\pi/4) \approx \text{cos}^2(k_hR-\pi/4) \approx 1/2$. In this approximation the proportionality factor between the Cooperon and the return probability $\hat{P}_{d,\omega}(\bm{r},\bm{r})$, see Eq.~(\ref{eq:pc11pd11f11}), is the same as in the normal state.

\section{\label{appendix:wl}Evaluation of the weak localization correction}

The energy-dependent return probability $\hat{P}_d(\bm{r},\bm{r})_{1,1}$, given in \autoref{eq:pdrrw0}, has different behaviors below and above $E_*$, see the definition of $\tau_{\text{min}}$ in \autoref{eq:taumin}. Accordingly, the energy integral for the WL correction to the thermal conductivity, Eq.~(\ref{eq:kc}), is split into two parts,
\begin{equation}
K_c=-\frac{1}{8\pi^2 k_BT^2}(I_1+I_2),
    \label{eq:kcappendix}
\end{equation}
which in two dimensions are explicitly
\begin{equation}
I_{\text{1}}=\int_\Delta^{E_*} dE \frac{E^2}{\cosh^2\left(\frac{E}{2k_BT}\right)}\ln\left[\frac{\tau_\phi (E^2-\Delta^2)}{\Delta}\right]
    \label{eq:i1}
\end{equation}
and
\begin{equation}
I_{\text{2}}=\int_{E_*}^\infty dE \frac{E^2}{\cosh^2\left(\frac{E}{2k_BT}\right)}\ln\left(\frac{\tau_\phi \sqrt{E^2-\Delta^2}}{\tau_e E}\right).
    \label{eq:i2}
\end{equation}
Below we consider two situations: energy-independent phase-coherence time $\tau_\phi$, and energy-independent phase-coherence length $L_\phi=\sqrt{D_s \tau_\phi}$. These two scenarios are equivalent in the normal state, but yield different results in the superconducting one.
We note that, strictly speaking, the lower integration limit of $I_1$
is not $\Delta$ but, in the diffusion approximation,
the quantity $\Delta_*$ defined by requiring that, for the left hand side of Eq.~(\ref{eq:pdtint}) to be non-zero, $\tau_\phi>\tau_\mathrm{min}$. For energy-independent phase time, under the usual assumption that $\tau_\phi \gg \tau_e$ (needed for the general applicability of the diffusive approximation~\cite{CCS}), we find for $\Delta_*$ the equation $\Delta_*^2=\Delta(1/\tau_\phi +\Delta)$; thus, for $\tau_\phi \gg 1/\Delta$, we have $\Delta_* \approx \Delta$, an approximation that is valid for temperature not too close to absolute zero, $k_BT \gg 1/\tau_\phi$ (at lower temperatures, the WL correction is, with logarithmic accuracy, absent, since the modes with energy between $\Delta$ and $\Delta^*$ are not diffusive).
The same approximation is valid in the case of energy-independent phase length (assumed to be long compared to the mean free path $l_e$) under the condition $L_\phi \gg \xi$, with $\xi = \sqrt{l_e \xi_\Delta}$, where $\xi_\Delta= v_F/\Delta$ is the BCS coherence length for a clean superconductor.

\subsection{Energy-independent \texorpdfstring{$\tau_{\phi}$}{tauphi} %
}

It is convenient to rewrite $I_1+I_2=I_{\text{n}}+I_{\varepsilon_*}+I_3$ with
\begin{eqnarray}
    I_{\text{n}}&=&\int^{\infty}_{\Delta} dE \frac{E^2}{\cosh^2\left(\frac{E}{2k_BT}\right)}\ln\left(\frac{\tau_\phi}{\tau_e}\right),
    \label{eq:inm}
    \\
    I_{\varepsilon_*}&=&\int^{E_*}_{\Delta} dE \frac{E^2}{\cosh^2\left(\frac{E}{2k_BT}\right)}\ln\left(\frac{E\sqrt{E^2-\Delta^2}}{E_*\sqrt{E_*^2-\Delta^2}}\right), \qquad
    \label{eq:ie*}
    \\
    I_3&=&\int^{\infty}_{\Delta} dE \frac{E^2}{\cosh^2\left(\frac{E}{2k_BT}\right)}\ln\left(\frac{\sqrt{E^2-\Delta^2}}{E}\right),
    \label{eq:i3}
\end{eqnarray}
where we have used the identity
\begin{equation}
\tau_e=\frac{\Delta}{E_*(E_*^2-\Delta^2)^{1/2}}.
    \label{eq:taueestar}
\end{equation}
which follows from the definition of $E_*$, see Eq.~(\ref{eq:varepsilonstar}).

The integral in \autoref{eq:inm} is defined such that its contribution to the relative correction to the thermal conductivity
$K_c/K_0$ coincides with that in the normal state, see
\autoref{eq:nmratio}. The other two integrals are then
responsible for the temperature-dependent deviations from the normal state expression.
We compute $I_3$ and $I_{\varepsilon_*}$ for different temperature regimes with logarithmic accuracy; note that only $I_{\varepsilon_*}$ depends on the disorder strength. 
We first consider the low-temperature regime $T \ll T_{\Delta}$ for both the dirty and the clean case, and later the high-temperature regime $T \gtrsim T_{\Delta}$.

\subsubsection{Low-temperature regime}

In the low-temperature regime, since we have $k_BT\ll\Delta$ the hyperbolic cosine can then be approximated as $1/\cosh^2(E/2k_BT)\approx 4 e^{-E/k_BT}$.
Introducing the dimensionless integration variable $\alpha=(E-\Delta)/k_BT$ and keeping only the leading term in the small parameter $k_BT/\Delta$, we find
\begin{equation}
    I_3\approx
    \frac{C}{2}\int_0^\infty \!d\alpha \, e^{-\alpha}
    \ln \left(\frac{2k_BT}{\Delta}\alpha\right) = \frac{C}{2}\ln\left(2e^{-\gamma_E}\frac{k_BT}{\Delta}\right)
    \label{eq:i3lt}
\end{equation}
with $C=4 k_BT \Delta^2 e^{-\Delta/k_BT}$ and $\gamma_E\simeq 0.5772\ldots$ the Euler-Mascheroni constant.

For the integral $I_{\varepsilon_*}$ we can proceed with the same approximation for the hyperbolic cosine and the same change of integration variable to get
\begin{equation}
    I_{\varepsilon_*}\approx
    \frac{C}{2}\int_0^{\alpha_*}\!d\alpha \, e^{-\alpha} \ln \frac{\alpha}{\left(1+\frac{k_BT}{\Delta}\alpha_*\right)^2\left(1+\frac{k_BT}{2\Delta}\alpha_*\right)\alpha_*}
\end{equation}
where $\alpha_* = \varepsilon_*/k_BT$. We must now treat separately the disordered ($\tau_e \Delta \ll 1$) and clean ($\tau_e \Delta \gg 1$) cases. In the disordered case we have $\alpha_* \gg \Delta/k_BT \gg 1$ and we obtain
\begin{equation}\label{eq:ie*ltd}
     I_{\varepsilon_*}\approx
    \frac{C}{2}\int_0\!d\alpha \, e^{-\alpha} \ln \frac{2\Delta^3 \alpha}{(k_BT)^3\alpha_*^4} = \frac{C}{2}\ln\left(\frac{2e^{-\gamma_E}\Delta^3k_BT}{\varepsilon_*^4}\right)
\end{equation}
The sum of Eqs.~(\ref{eq:i3lt}) and (\ref{eq:ie*ltd}) leads to the last term in the top line of Eq.~(\ref{eq:dirtylt}).

In the clean case, since $\alpha_* k_BT/\Delta \ll 1$, the integral simplifies to
\begin{equation}\label{eq:iveclht}
     I_{\varepsilon_*}\approx
    \frac{C}{2}\int_0^{\alpha_*}\!d\alpha \, e^{-\alpha} \ln \frac{\alpha}{\alpha_*} \, .
\end{equation}
At very low temperatures such that $k_B T \ll \varepsilon_*$ we can extend the upper integration limit to infinity and thus find a logarithmic contribution of the form $I_{\varepsilon_*}\approx C \ln \left(e^{-\gamma_E}k_BT/\varepsilon_*\right)/2$; we also note here that for this contribution to be present the condition $\tau_\phi \gg 1/\Delta$ mentioned above is not sufficient, and a more stringent one ($\tau_\phi \gg \tau_e^2 \Delta$), obtained from demanding $E_* \gg \Delta_*$, is needed. At intermediate temperatures $\varepsilon_* \ll k_BT \ll k_BT_\Delta$, on the other hand, there is no logarithmic contribution from $I_{\varepsilon_*}$ and hence the last term in the top line of Eq.~(\ref{eq:cleanlt}) is determined solely by Eq.~(\ref{eq:i3lt}).

\subsubsection{High-temperature regime}

In the high-temperature regime $T \gtrsim T_{\Delta}$, we can approximate $k_B T \gg \Delta$. The integral $I_3 \sim k_BT\Delta^2$ has then no logarithmic parameter dependence and can be  neglected in comparison to $I_{\text{n}}\sim (k_BT)^3\ln\left(\tau_\phi/\tau_e\right)$. For $I_{\varepsilon_*}$ we must again consider the various regimes separately. However, for $k_BT$ large compared to $E_*$ (which is always true in the clean case at high temperatures, while it would require $T$ in the narrow range between $T_*$ and $T_c$ for the dirty case), we can approximate the hyperbolic cosine with unity; then $I_{\varepsilon_*}$ becomes independent of temperature and displays no logarithmic parameter dependence; thus, as $I_3$ above, $I_{\varepsilon_*}$ can be neglected in comparison to $I_\text{n}$ and we arrive at the result in the top line of Eq.~(\ref{eq:nmratio}).

We are left with the dirty case in the regime $T_\Delta \lesssim T \lesssim T_*$. Then $\Delta$ is small compared to both $E_*$ and the typical energy $E \sim T$, so that we can write
\begin{equation}
I_{\varepsilon_*} \approx \int^{E_*}_{\Delta} \! dE\, \frac{E^2}{\cosh^2\left(\frac{E}{2k_BT}\right)}\, 2\ln\left(\frac{E}{E_*}\right)
\end{equation}
which, with logarithmic accuracy, is
\begin{equation}
I_{\varepsilon_*}=I_{K_0}
2\ln\left(\frac{k_BT}{\varepsilon_*}\right)
\end{equation}
with
\begin{equation}
    I_{K_0}=\int_{\Delta}^{\infty}  \! dE  \, \frac{E^2}{\cosh^2\left(\frac{E}{2 k_BT}\right)}.
    \label{eq:ik0}
\end{equation}
Since we can also write $I_\mathrm{n} = I_{K_0} \ln \left(\tau_\phi/\tau_e\right)$, the sum $I_\mathrm{n} + I_{\varepsilon_*}$ leads to the top line in Eq.~(\ref{eq:dirtyht}).

\subsection{Energy-independent \texorpdfstring{$L_{\phi}$}{Lphi} %
}

In the previous subsection, we assumed the phase-coherence time to be independent of energy. Since the group velocity $v_g$ [Eq.~(\ref{eq:taus})] in a superconductor and hence the diffusion constant $D_s$ [Eq.~(\ref{eq:ds})] are energy dependent, such a choice for the phase-coherence time leads to an energy-dependent phase-coherence length.
As an alternative scenario, we consider here a constant phase-coherence length, expressed in terms of the dephasing time and diffusion constant as $L_\phi=\sqrt{D_s\tau_\phi}$.
This choice now leads to an energy dependent phase-coherence time $\tau_\phi= L_\phi^2/ l_e v_g$. We substitute this expression for $\tau_\phi$ together with $\tau_e=l_e/v_F$ in \autoref{eq:i1} and \autoref{eq:i2} to rewrite the integrals in terms of length rather than time scales. We obtain $I_1+I_2=I_\text{n}+I_{\varepsilon_*}$, with%
\begin{equation}
    I_{\text{n}}=\int^{\infty}_{\Delta_*} \! dE \, \frac{E^2}{\cosh^2\left(\frac{E}{2k_BT}\right)}\ln\left(\frac{L_\phi^2}{l_e^2}\right)
    \label{eq:inmfl}
\end{equation}%
and $I_{\varepsilon_*}$ as defined in Eq.~(\ref{eq:ie*}).
The expressions for the different regimes can then be easily obtained using the results for $I_{\varepsilon_*}$ in the preceding part of the appendix. Here we only note that the condition for the presence of the $I_{\varepsilon_*}$ contribution in the clean case for the lowest temperature regime $T \ll T_*$ [see discussion after Eq.~(\ref{eq:iveclht})] can be written as $L_{\phi}\gg l_e$.

 \section{\label{appendix:so}Weak anti-localization: Spin-orbit scattering}

In this appendix, we study weak anti-localization~\cite{HLN} in the presence of spin-orbit scattering in disordered $s$-wave superconductors.
To properly account for spin, we now define the Nambu vector as [cf. Eq.~(\ref{eq:nambubasis})]
\begin{equation}
\bm{\Psi}_{\bm{k}}= \begin{pmatrix} c_{\bm{k}\uparrow} \\ c_{\bm{k}\downarrow} \\ T\begin{bmatrix} c_{\bm{k}\uparrow} \\ c_{\bm{k}\downarrow}  \end{bmatrix} \end{pmatrix} = \begin{pmatrix}c_{\bm{k}\uparrow} \\ c_{\bm{k}\downarrow} \\   c^\dag_{-\bm{k} \downarrow} \\ -c^\dag_{-\bm{k} \uparrow} \end{pmatrix}.
\label{eq:nambuspin}
\end{equation}%
The full Nambu space is then the product between the space spanned by $\{\ket{\text{e}},\ket{\text{h}}\}$ (the basis used in the main text) and the spin space spanned by $\{\ket{\uparrow},\ket{\downarrow}\}$; the Pauli matrices $\tau_i$ and $\sigma_i$ act respectively on these two subspaces. The spin-orbit scattering can be expressed as an additional term in the Hamiltonian in the form~\cite{AM}
\begin{equation}
\hat{H}_{\alpha\alpha'}^{\text{so}}(\bm{k},\bm{k}')=i V^{\text{so}}\bm{\sigma}_{\alpha\alpha'}\cdot(\bm{u_k}\times\bm{u_{k'}}) \otimes \tau_3,
\label{eq:Hso}
\end{equation}%
where $V^{\text{so}}$ is the strength of the spin-orbit scattering potential, $\bm{u_k}=\bm{k}/k$, the components of the operator $\bm{\sigma}$ are the Pauli matrices $\{\sigma_x,\sigma_y,\sigma_z\}$ and $\bm{\sigma}_{\alpha\alpha'}=\bra{\alpha'}\bm{\sigma}\ket{\alpha}$ with $\alpha,\alpha' \in \{\uparrow,\downarrow\}$.
The full disorder potential now takes the form $\hat{V}_{\alpha\alpha'}(\bm{k},\bm{k}')=V_{\alpha \alpha'}(\bm{k},\bm{k}')\otimes \tau_3$ with
\begin{equation}
V_{\alpha\alpha'}(\bm{k},\bm{k}') =V_0\delta_{\alpha\alpha'}+i V^{\text{so}}\bm{\sigma}_{\alpha\alpha'}\cdot (\bm{u_k}\times\bm{u_{k'}}).
\label{eq:vso}
\end{equation}%
This leads to a new disorder parameter $\gamma_{\text{tot}}=\langle \overline{|V_{\alpha \alpha'}(\bm{k},\bm{k}')|^2 }\rangle_{\bm{k}'}=\gamma_e + \gamma_{\text{so}}$, with $\gamma_{\text{so}}=1/2\pi \rho_0 \tau_{\text{so}}$,  $\tau_{\text{so}}=l_{\text{so}}/v_F$ and where $\gamma_e$ has been defined at the end of \autoref{sec:model}.

\begin{figure}[!tb]
  \includegraphics[width=\linewidth]{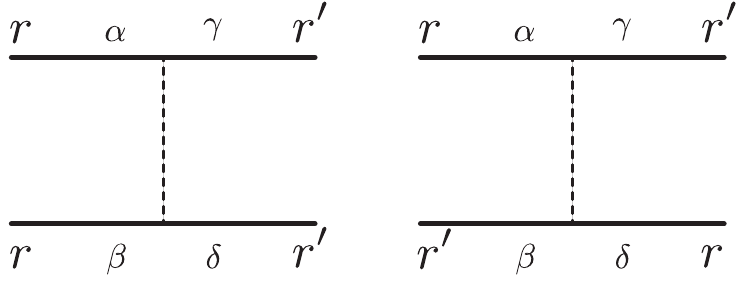}
    \caption{Elementary vertex with spin-orbit impurity scattering for the diffuson (left) and Cooperon (right).}
    \label{fig:spin}
\end{figure}

The disorder-averaged superconducting Green's function can be generalized to the full Nambu space as
\begin{equation}
    \overline{\hat{G}_E^{R,A}}=\begin{pmatrix} \overline{G_E^{R,A}} & \overline{ F_E^{R,A}}\\  \overline{\bar{F}_E^{R,A}} &  \overline{\bar{G}_E^{R,A}} \end{pmatrix} \otimes \sigma_{0},
    \label{eq:greennambuspin}
\end{equation}
and the diffuson and the Cooperon can be calculated following a procedure similar to the one used in \autoref{sec:disorder}. We define
\begin{equation}
\hat{p}_{{d,\omega}}^{{\text{so}}}(\bm{r},\bm{r'})=\langle \hat{P}_{{0}}^{{\text{so}}}\rangle_{\bm{r}}\hat{\Gamma}_{\omega}^{{\text{so}}}(\bm{r},\bm{r}') \langle \hat{P}_{{0}}^{{\text{so}}}\rangle_{\bm{r}},
\label{eq:pdso}
\end{equation}%
\begin{equation}
\hat{p}_{{c,\omega}}^{{\text{so}}}(\bm{r},\bm{r'})=\hat{F}^{\text{so}}(\bm{R})\hat{\Gamma}_{{c,\omega}}^{{\text{so}}}(\bm{r},\bm{r}) \hat{F}^{{\text{so}}}(\bm{R}),
\label{eq:pcso}
\end{equation}%
which generalize \autoref{eq:pddefp0gp0} and \autoref{eq:pcdeffgf}, respectively.
We use lower-case $p$s to emphasize that not all elements of these matrices correspond to diffusons and Cooperons, as we will later see. 
The terms that do not take collisions into account, \textit{i.e} $\langle \hat{P}_0^{\text{so}} \rangle_{\bm{r}}$ and $\hat{F}^{\text{so}}(\bm{R})$,
are related to those in the absence of spin-orbit scattering by $\langle \hat{P}_{{0}}^{{\text{so}}} \rangle_{\bm{r}}=\langle \hat{P}_{{0}} \rangle_{\bm{r}}\otimes \sigma_0 \otimes \sigma_0$ and $\hat{F}^{{\text{so}}}(\bm{R})=\hat{F}(\bm{R})\otimes \sigma_0 \otimes \sigma_0$; here $\langle \hat{P}_0 \rangle_{\bm{r}}$ and $\hat{F}(\bm{R})$ are as those defined in \autoref{eq:p0r} and \autoref{eq:Frdeffull}, respectively, but with $\gamma_{\text{tot}}$ replacing $\gamma_e$.
The equations followed by the structure factors are now given by
\begin{equation}
    \hat{M}_{\omega}^{\text{so}}(\bm{r}) \hat{\Gamma}_{{\omega}}^{{\text{so}}}(\bm{r},\bm{r}')=\gamma_e \delta^{(d)}(\bm{r}'-\bm{r})
    \label{eq:gamsoeq}
\end{equation}
\begin{equation}
    \hat{M}_{c,\omega}^{\text{so}}(\bm{r}) \hat{\Gamma}_{{c,\omega}}^{{\text{so}}}(\bm{r},\bm{r}')=\gamma_e \delta^{(d)}(\bm{r}'-\bm{r}).
    \label{eq:gamcsoeq}
\end{equation}
The diffusion matrices $\hat{M}_{\omega}^{\text{so}}(\bm{r})$ and $\hat{M}_{c,\omega}^{\text{so}}(\bm{r})$ are each defined by an equation similar to \autoref{eq:mdef}, but substituting $\langle \hat{P}_{{0}} \rangle_{\bm{r}}$ by $\langle \hat{P}_{{0}}^{{\text{so}}} \rangle_{\bm{r}}$ and $\hat{U}_v$ by the potential matrices $\hat{U}_v^{\text{so}}$ and $\hat{U}_{c,v}^{\text{so}}$.
The potential matrices are no longer equivalent for the diffuson and the Cooperon due to the different spin and momenta relations between the retarded and advanced Green's functions in the two cases. They are given by
\begin{equation}
 \hat{U}_v^{\text{so}}=\hat{U}_v \otimes \hat{u}^{\text{so}}
  \label{eq:uvso}
\end{equation}%
\begin{equation}
 \hat{U}_{c,v}^{\text{so}}=\hat{U}_v \otimes \hat{u}_c^{\text{so}},
  \label{eq:uvcso}
\end{equation}%
with the (normal metal~\cite{AM}) matrices $\hat{u}^{\text{so}}$ and $\hat{u}_c^{\text{so}}$ given in the basis $\{ \ket{ \uparrow \uparrow },\ket{ \uparrow \downarrow },\ket{ \downarrow  \uparrow },\ket{ \downarrow  \downarrow  } \}$ by
\begin{equation}
\hat{u}^{\text{so}} = \begin{pmatrix} 1 & 0 & 0 & 0 \\ 0 & 1 & 0 & 0 \\ 0 & 0 & 1 & 0 \\ 0 & 0 & 0 & 1 \end{pmatrix} +\frac{\gamma_{\text{so}}}{3\gamma_e}\begin{pmatrix} 1 & 0 & 0 & 2 \\ 0 & -1 & 0 & 0 \\ 0 & 0 & -1 & 0 \\ 2 & 0 & 0 & 1 \end{pmatrix},
\label{eq:uso}
\end{equation}%
\begin{equation}
\hat{u}_c^{\text{so}} = \begin{pmatrix} 1 & 0 & 0 & 0 \\ 0 & 1 & 0 & 0 \\ 0 & 0 & 1 & 0 \\ 0 & 0 & 0 & 1 \end{pmatrix} -\frac{\gamma_{\text{so}}}{3\gamma_e}\begin{pmatrix} 1 & 0 & 0 & 0 \\ 0 & -1 & 2 & 0 \\ 0 & 2 & -1 & 0 \\ 0 & 0 & 0 & 1 \end{pmatrix}.
\label{eq:ucso}
\end{equation}%
Here, each element $\bra{\gamma \delta} \hat{u}^{\text{so}} \ket{\alpha \beta } $ and $\bra{\gamma \delta} \hat{u}_c^{\text{so}} \ket{\alpha \beta } $ with $\alpha,\beta,\gamma,\delta \in \{\uparrow,\downarrow\}$, relates the spins of the Green's functions before and after interacting with an impurity, as depicted in \autoref{fig:spin}.
After obtaining $\hat{\Gamma}_{\omega}^{\text{so}}(\bm{r},\bm{r}')$ and $\hat{\Gamma}_{c,\omega}^{\text{so}}(\bm{r},\bm{r})$ from \autoref{eq:gamsoeq} and \autoref{eq:gamcsoeq}, the matrices $\hat{p}_{d,\omega}^{\text{so}}(\bm{r},\bm{r}')$ and $\hat{p}_{c,\omega}^{\text{so}}(\bm{r},\bm{r}')$ can be calculated using \autoref{eq:pdso} and \autoref{eq:pcso}.

Not all terms in $\hat{p}_{d,\omega}^{\text{so}}(\bm{r},\bm{r}')$ and $\hat{p}_{c,\omega}^{\text{so}}(\bm{r},\bm{r}')$ represent diffusons or Cooperons. The trajectories represented by the retarded and advanced Green's functions in the diffuson and the Cooperon are not independent and their spin configurations are related. The diffuson, for instance, is composed by a time reversed pair of trajectories; this implies that $\alpha=\beta$ and $\gamma=\delta$. We can obtain the diffuson by summing over the final spin configuration while taking this constraint into account. In this way we recover a 4$\times$4 matrix in Nambu space, similar to $\hat{P}_{d,\omega}$ of \autoref{sec:disorder}, where each element now accounts for the probability of propagation with and without spin flip. The diffuson for a particle with initial spin $\alpha$ is given by
\begin{equation}
\bra{i',j'}\hat{P}_{d,\omega}^{\text{so}}(\bm{r},\bm{r}')\ket{i,j}=\sum_{\beta}\bra{i'_\beta,j'_\beta}\hat{p}_{d,\omega}^{\text{so}}(\bm{r},\bm{r}')\ket{i_\alpha,j_\alpha},
\label{eq:diffspin}
\end{equation}%
where $\ket{i},\ket{j} \in \{\ket{\text{e}},\ket{\text{h}}\}$ and   $\ket{i_\alpha},\ket{j_\alpha} \in \{\ket{\text{e}}\otimes\ket{\alpha},\ket{\text{h}}\otimes \ket{\bar{\alpha}}\}$ with $\alpha\in \{\uparrow,\downarrow\}$ and $\bar{\alpha}\neq \alpha$.
The Cooperon also accounts for the probability of propagation with and without spin flip; however, the conditions on the spins are different since the advanced Green's function (lower line in \autoref{fig:spin}) now covers the trajectory in the opposite direction. It is now necessary that $\alpha=\delta$ and $\gamma=\beta$, and the Cooperon contribution for a particle with initial spin $\alpha$ is given by
\begin{equation}
\bra{i',j'}\hat{P}_{c,\omega}^{\text{so}}(\bm{r},\bm{r}')\ket{i,j}=\sum_{\beta}\bra{i'_\alpha,j'_\beta}\hat{p}_{c,\omega}^{\text{so}}(\bm{r},\bm{r}')\ket{i_\beta,j_\alpha}.
\label{eq:coopspin}
\end{equation}%

Direct calculation (cf. Ref.~\cite{AM}) shows that spin-orbit scattering does not affect the diffuson, $\hat{P}_{d,\omega}^{\text{so}}(\bm{r},\bm{r}')=\hat{P}_{d,\omega}(\bm{r},\bm{r}')$, while the Cooperon is now qualitatively different, with
\begin{equation}
\hat{P}_{{c,\omega}}^{{\text{so}}}(\bm{r},\bm{r'})=-\frac{1}{2}\hat{P}_{c,\omega}(\bm{r},\bm{r'}).
\label{eq:pcsopc}
\end{equation}%
As a consequence, in the presence of spin-orbit scattering the quantum correction to the thermal conductivity is
\begin{equation}
\frac{K_{c}^{{\text{so}}}}{K_0^{\text{so}}}=-\frac{1}{2}\frac{K_c}{K_0},
\label{eq:kcso}
\end{equation}%
where $K_c/K_0$ is the correction calculated in \autoref{sec:thermalconductivity}. This correction, known as weak-anti-localization (WAL) effect, increases the total thermal conductivity
and is due to destructive interference between self-crossing paths.

\bibliography{refs}

\end{document}